# Quantized Thermoelectric Hall Effect Induces Giant Power Factor in a Topological Semimetal

Fei Han[1*†], Nina Andrejevic[2†], Thanh Nguyen[1†], Vladyslav Kozii[3†], Quynh T. Nguyen[1,3], Tom Hogan[4], Zhiwei Ding[2], Ricardo Pablo-Pedro[1], Shreya Parjan[5], Brian Skinner[3], Ahmet Alatas[6], Ercan Alp[6], Songxue Chi[7], Jaime Fernandez-Baca[7], Shengxi Huang[8], Liang Fu[3*], Mingda Li[1*]

[1]Department of Nuclear Science and Engineering, Massachusetts Institute of Technology, Cambridge, MA 02139, USA

[2]Department of Materials Science and Engineering, Massachusetts Institute of Technology, Cambridge, MA 02139, USA

[3]Department of Physics, Massachusetts Institute of Technology, Cambridge, MA 02139, USA

[4]Quantum Design Inc, San Diego, CA 92121

[5]Department of Physics, Wellesley College, 106 Central St, Wellesley, MA 02481, USA

[6]Advanced Photon Source, Argonne National Laboratory, Lemont, IL 60439, USA

[7]Neutron Scattering Division, Oak Ridge National Laboratory, Oak Ridge, TN, 37831, USA

[8]Department of Electrical Engineering, The Pennsylvania State University, State College, PA 16802, USA

## Abstract

Thermoelectrics are promising by directly generating electricity from waste heat. However, (sub-)room-temperature thermoelectrics have been a long-standing challenge due to vanishing electronic entropy at low temperatures. Topological materials offer a new avenue for energy harvesting applications. Recent theories predicted that topological semimetals at the quantum limit can lead to a large, non-saturating thermopower and a quantized thermoelectric Hall conductivity approaching a universal value. Here, we experimentally demonstrate the non-saturating thermopower and quantized thermoelectric Hall effect in the topological Weyl semimetal (WSM) tantalum phosphide (TaP). An ultrahigh longitudinal thermopower $S_{xx} = 1.1 \times 10^3 \mu V K^{-1}$ and giant power factor $\sim 525 \mu W cm^{-1} K^{-2}$ are observed at $\sim$40K, which is largely attributed to the quantized thermoelectric Hall effect. Our work highlights the unique quantized thermoelectric Hall effect realized in a WSM toward low-temperature energy harvesting applications.

†These authors contribute equally to this work.
*Corresponding authors: hanfei@mit.edu; liangfu@mit.edu; mingda@mit.edu.

## Introduction

Over two-thirds of global energy production is rejected as waste heat. Thermoelectrics are attractive by directly converting waste heat into electricity without moving parts. The efficiency of thermoelectric energy conversion is an increasing function of a dimensionless quantity $zT = \sigma S^2 T / \kappa$, where $\sigma$, $S$, and $\kappa$ denote the electrical conductivity, thermopower, and total thermal conductivity, respectively[1]. Conventional thermoelectrics largely focus on tuning the thermal and electrical conductivities. Many efforts, such as lowering dimensionality[2], microstructuring[3,4] and nanostructuring[5,6], share the same principle: By increasing the scattering of major heat carriers of long mean-free-path phonons without affecting the short mean-free-path electrons, a level of independent tunability between electrical conductivity σ and thermal conductivity κ can be achieved, such as the phonon-glass electron-crystal state[7]. However, less attention was paid to improve the thermopower S, even though the $S^2$ dependence in $zT$ makes such improvement appealing. Moreover, thermopower S is proportional to the entropy per carrier and is therefore suppressed at reduced temperature[8]. For this reason, current thermoelectrics are generally effective only at elevated temperatures, and there is a pressing need for thermoelectrics that work efficiently at room-temperature and below. Filling this need requires new materials that can exhibit large electronic entropy at low temperatures while maintaining significant electrical conductivity.

One approach to creating large electronic entropy is bandstructure engineering through low carrier density, partially-filled carrier pockets[9]; a similar principle has also been applied to semimetals, such as Bi[10], graphite[11], and most recently WSMs, to explore large entropy at low carrier density[12-14]. However, the electrical conductivity is thereby reduced. Magnetic field offers an additional incentive to dramatically increase the entropy since the linear field-dependence of the density of states (DOS) enables unbounded, macroscopic number of states in each Landau level (LL), yet in conventional thermoelectrics, charge carriers will be localized at high $B$-field due to the cyclotron motion, still resulting in low conductivity. Consequently, increasing power factor ($\equiv \sigma S^2$) creates a significant challenge as it requires optimization of both $\sigma$ and $S$ under conflicting conditions.

The recent development of topological materials[15,16], including topological WSMs[17], offers a new pathway to surpass conventional thermoelectrics that relies on the topological protection of electronic states[18,19]. It is particularly worthy to note that the WSM system has a unique $n=0$ LL, which has a highly unusual, energy-independent DOS $g(n=0) = N_f Be / 4\pi^2 \hbar^2 v_F$ increasing linearly with $B$, and therefore can create huge electronic entropy. More importantly, the system remains gapless under high field thanks to the topological nature of Weyl nodes. Consequently, recent theories predicted a non-saturating thermopower[20] and quantized thermoelectric Hall conductivity at the quantum limit[21], where electrons and holes contribute additively to high thermoelectric performance without experiencing localization.

In this work, we carry out high-precision thermoelectric measurements using a centimeter-sized crystal WSM TaP (Figure 1a and b, and Supplementary Information I, II). The Fermi level is fine-tuned through the synthesis procedure to approach the $n=0$ LL near the W2 Weyl node (Figure 1g). In this system, giant, non-saturating longitudinal thermopower $S_{xx}$ is observed, which exhibits linear dependence with $B$- field without saturation. Additionally, the signature of the quantized thermoelectric Hall conductivity is observed, where the low-temperature, high-field thermoelectric Hall conductivity $\alpha_{xy} \equiv [\rho^{-1}S]_{xy}$ approaches a universal curve determined by number of fermion flavors, Fermi velocity, and universal constants. Moreover, evidence of Wiedemann-Franz law violation further indicates a breakdown of quasiparticle behaviors. Our work leverages the effects of topology to overcome challenges for low-temperature thermoelectric energy harvesting from a power factor perspective.

## Results

**Quantum oscillations.** We first present the longitudinal magnetoresistance (MR) data, where Giant MR was observed. At $T$<25K, the $MR \equiv \left(\rho_{xx}(B) - \rho_{xx}(0\mathrm{T})\right) \big/ \rho_{xx}(0\mathrm{T})$ exceeds $10^5$% (Figure 1c). This is a signature of electron-hole compensation, which is further confirmed by the two-band model fitting of conductivity, with $n_e = 2.39 \times 10^{19}\,\mathrm{cm}^{-3}$ and $n_h = 2.35 \times 10^{19}\,\mathrm{cm}^{-3}$ at $T$=2.5K, along with a high mobility of $\sim 1 \times 10^5\,\mathrm{cm}^2\mathrm{V}^{-1}\mathrm{s}^{-1}$ (Supplementary Information III). The background-subtracted MR, denoted $\Delta MR$, exhibits Shubnikov-de Haas (SdH) oscillations, which are plotted against $1/B$ to determine oscillation frequencies (Figure 1d). The Fourier transform of $\Delta MR$ shows two small carrier pockets with low frequency $F_\alpha = 4\mathrm{T}$ and $F_\beta = 18\mathrm{T}$ among four pockets (Supplementary Information IV, Figure 1e). The LL fan diagram analysis indicates the two small pockets are at $n$=2 LL and $n$=0 LL, respectively (Supplementary Information V, Figure 1f). The intersections of the linear LL index plots (-0.037 for $n$=0 LL and +0.065 for $n$=2) lying between -1/8 to +1/8 indicates that the two pockets are both topologically nontrivial[22,23], from which we attribute the *n=2* LL to the electron pocket of the W1 Weyl node, and the *n=0* LL to the hole pocket of the W2 Weyl node (Figure 1g). Moreover, we see that the W2 and W1 pockets enter the quantum limit at $B \sim 3.8\mathrm{T}$ and 16T, respectively. There is an alternate way to infer LL. The Weyl fermion dispersion of the $n^{th}$ LL at $k_z = 0$ is given by $E_n = \text{sgn}(n)v_F\sqrt{2e\hbar B|n|}$ , while the oscillation frequency $F$ satisfies $F = E_F^2/2e\hbar v_F^2$ . When $E_n \sim E_F$, we have $F \sim B|n|$. This leads to an agreement between $n$=2 LL and the measured $F_\beta = 18\text{T}$ at $B \sim 9\text{T}$. For $F_\alpha$, the low frequency 4T suggests an extremely small Fermi surface. Since the spacing between $n$=1 and $n$=0 LLs is given by $E_1 - E_0 = v_F\sqrt{2e\hbar B} = E_F\sqrt{B/F}$, the condition to reach the $n$=0 LL quantum limit for W2 pocket is met at $B > F_\alpha = 4\text{T}$. This value agrees well with the above LL index analysis.

**Non-saturating thermopower and giant power factor.** Having determined the carrier characteristics, we carried out thermoelectric measurements using a diagonal offset geometry (Figure 2a), where the electrical and thermal transport along both the longitudinal and transverse directions can be acquired together by flipping the field polarity (Supplementary Information VI, which also contains the phase relations between various thermoelectric quantities). The longitudinal thermopower $S_{xx}$ is shown in Figure 2b, where $S_{xx}$ increases over 200-fold compared to its zero-field value, and reaches a giant magnitude of $1.07 \times 10^3\,\mu\text{VK}^{-1}$ without sign of saturation at $B = 9\text{T}$ and $T = 40\text{K}$. One prominent feature is that $S_{xx}$ develops a double-peak behavior, which may be attributed to the two types of Weyl nodes: The higher carrier mobility and lower carrier density at the W2 node leads to reduced phonon scattering, and thus the high $S_{xx}$ can persist to higher temperatures. Quantitatively, it has been predicted that for the $n$=0 chiral LL of Weyl electrons, $S_{xx}$ obeys a simple formula[20]:

$$S_{xx} = \frac{k_B^2}{\hbar^2}\frac{N_f}{12}\frac{TB}{v_F^{\text{eff}}(n_h - n_e)} \qquad (1)$$

where $N_f$ is the degeneracy of the Weyl nodes, $n_h - n_e$ is the net carrier density, and $v_F^{\text{eff}}$ is an effective Fermi velocity. Since TaP has two sets of Weyl nodes with different velocities and energies, in this work we introduce $v_F^{\text{eff}}$ as an effective parameter capturing an average Fermi velocity of the system.

The linearity of $S_{xx}$ with $T$ and $B$ is shown in Figures 2c and 2d, respectively. It is noteworthy that Eq. (1) is in quantitative agreement with our result if we adopt the fitted value of the $v_F^{\text{eff}}$ using Eq. (3) and Eq. (S14), described in greater detail in the following section. Such quantitative agreement is valid across all fields and up to ~40K and is a measure of the success of the effective model (Figure 2e). Moreover, a giant longitudinal power factor $\equiv S_{xx}^2/\rho_{xx}$ up to 525μWcm⁻¹K⁻² is achieved due to the large entropy generated by the linearly-dispersive bands at quantizing magnetic fields, while a low $\rho_{xx}$ is maintained due to the protection of the gapless $n=0$ LL, evading the typical fate of carrier cyclotron motion and localization under fields[20,21]. In fact, this value is an order of magnitude higher than peak values of promising thermoelectrics (e.g., 10μWcm⁻¹K⁻² for SnSe at ~800K[24]) and two orders of magnitude higher than non-topological semimetals[10,11], which can achieve high thermopower at high magnetic fields with linear-dispersive bands, but cannot simultaneously maintain a low magneto-electrical resistivity.

**Quantized thermoelectric Hall effect.** Regarding the transverse properties, we see that the transverse thermopower $S_{yx}$ exhibits a plateau with increasing $B$-field, which is known to originate from the constant $k$-space volume as thermopower is a measure of occupational entropy in state space[12]. The thermoelectric Hall conductivity $\alpha_{xy} \equiv (S_{xx}\rho_{yx} - S_{yx}\rho_{xx})/(\rho_{xx}^2 + \rho_{yx}^2)$ is shown in Figure 3b, where in the low-temperature range, the flatness with respect to $B$-field starts to emerge. In particular, under the low-temperature $k_B T \quad E_F$ and high-field $B \gg E_F^2/\hbar e v_F^2$ limit, $\alpha_{xy}$ is predicted to approach the following universal value that is independent of $B$-field, disorder, carrier density, and even carrier type[21]:

$$\alpha_{xy,\text{ideal}} = \frac{\pi^2}{3} \frac{ek_B^2}{(2\pi\hbar)^2} \frac{N_f}{v_F^{\text{eff}}} T \tag{2}$$

The temperature dependence of $\alpha_{xy}$ is shown in Figure 3c, where we see that the linearity with $T$ holds up to $T \sim 10 K$. As a direct consequence, the $\alpha_{xy}/T$ curves converge to a single curve at high fields (Figure 3d and 3e), where an ideal value $\alpha_{xy,\text{ideal}}/T = 0.4 \, \text{A K}^{-2} \text{m}^{-1}$ is determined by evaluating Eq. (2) using $v_F^{\text{eff}} = 1.4 \times 10^4 \, \text{ms}^{-1}$, which is extracted by fitting a more general Eq. (3) at base temperature:

$$\alpha_{xy} = \frac{eN_f}{2\pi\hbar} \sum_{n=0}^{\infty} \int_0^{\infty} \frac{dk_z}{\pi} \left[ s\left( \frac{\varepsilon_n^0(k_z) - \mu}{k_B T} \right) + s\left( \frac{\varepsilon_n^0(k_z) + \mu}{k_B T} \right) \right] \tag{3}$$

in which $s$ is the electronic entropy function (Eq. S13). The magnitude $v_F^{\text{eff}}$ is comparable to the simple weighted average of projected Fermi velocity $v_{F,z}^{W1} = 0.77 \times 10^5 \, \text{ms}^{-1}$, $v_{F,z}^{W2} = 1.88 \times 10^5 \, \text{ms}^{-1}$ [25], which gives $\overline{v}_{F,z}^{\text{eff}} = 1.5 \times 10^5 \, \text{ms}^{-1}$, where the $z$-direction was chosen to coincide with the magnetic field direction. The fitted chemical potential $\mu$ is consistent with the electrical transport measurements, while the $v_F^{\text{eff}}$ is lower than the $v_F$ at W2[25]. This can be understood since carriers at W1 Weyl nodes at $n=2$ LL have yet to reach extreme quantum limit (Figures 1g, 3f and Supplementary Information VI and VII). For temperatures above 10K, scattering effects are significant and the dissipationless limit assumed in Eq. (3) is no longer valid; thus, for fits at $T \geq 10 K$, approximate forms of $\alpha_{xy}$ which include a finite scattering time were used (Eq. (S14) and Eq. (S16)). To corroborate the universal quantization behavior of $\alpha_{xy}/T$, we performed separate thermoelectric measurements up to $B=14 T$ at $T=2K$, 4K and 6K, where the collapse onto a single curve and a clearer plateau are observed (Supplementary Information VIII), in addition to giving $\alpha_{xy,\text{ideal}}/T = 0.37 \, \text{A K}^{-2} \text{m}^{-1}$, in quantitative agreement with the 9T data. Finally, to show that the quantized thermoelectric Hall coefficient $\alpha_{xy}$ drives the ultrahigh thermopower and giant power factor, we decompose $S_{xx}$ into its transverse ($-\alpha_{xy}\rho_{xy}$) and longitudinal ($\alpha_{xx}\rho_{xx}$) components, where we see that the transverse term $\alpha_{xy}\rho_{xy}$ contributes to ~90% of the longitudinal $S_{xx}$ (Figure 4a and Supplementary Information IX). The corresponding decomposed contributions to power factor $S_{xx}^2/\rho_{xx}$ is shown in Figure 4b.

In a nutshell, the quantized $\alpha_{xy}$, large non-saturating $S_{xx}$, and ultrahigh power factor $S_{xx}^2/\rho_{xx}$ all originate from the topological Weyl nodes, but with increasingly stringent manifestation conditions: the quantized $\alpha_{xy}$ comes directly from the gapless $n=0$ LL states of Weyl fermions; since $S_{xx} = -\alpha_{xy}\rho_{xy} + \alpha_{xx}\rho_{xx}$, $\rho_{xy}$ should increase at high fields to obtain non-saturating $S_{xx}$ with the field-independent $\alpha_{xy}$; only when the transverse components $-\alpha_{xy}\rho_{xy}$ dominate the $S_{xx}$ with moderate $\rho_{xx}$ can the power factor $S_{xx}^2/\rho_{xx}$ be enhanced – the gapless $n=0$ LL states can also contribute to reduce the $\rho_{xx}$.

**Breakdown of the Wiedemann-Franz Law.** Wiedemann-Franz (WF) law is a robust empirical law stating that the ratio between the electronic thermal conductivity $\kappa^e$ and electrical conductivity $\sigma$ are related by a universal Lorenz number:

$$L_0 \equiv \frac{\kappa^e}{\sigma T} = \frac{\pi^2}{3}\left(\frac{k_B}{e}\right)^2 = 2.44 \times 10^{-8}\,\mathrm{W\Omega K^{-2}}. \tag{4}$$

Recently, it has been reported that there is strong violation of the WF law in the 2D Dirac fluid of graphene[26] and WSM WP$_2$[27] due to collective electron hydrodynamics. Other behaviors of electrons, like quantum criticality[28] or quasiparticle breakdown[29,30], can also lead to the WF law violation. It is thus worthwhile to examine the validity of the WF law in the field-induced high-entropy state of TaP. To do so, it is crucial to properly separate $\kappa^e$ from the lattice thermal conductivity $\kappa^{ph}$. We adopt the following empirical relation by using the field-dependence of $\kappa^e$ [31]:

$$\kappa_{xx}(T,B) = \kappa_{xx}^{ph}(T) + \frac{\kappa_{xx}^e(T)}{1+\beta_e(T)B^m}. \tag{5}$$

where $\beta_e(T)$ is a measure of zero-field electron mean free path and $m$ is a factor related to the nature of scattering. Figure 5a demonstrates an example for such a separation procedure (Supplementary Information XI). Using this method, we see that the extracted $\kappa^{ph}$ agrees well with the computed value from ab initio calculations (Figure 5b and Supplementary Information XII), from which the phonon dispersions are also computed, and agree well with measured dispersion from inelastic scattering (Figure 5c and Supplementary Information X). All these agreements indicate the reliability of the separation process. The corresponding $\kappa^e$ and the $L_0$ is shown in Figure 5d and Figure 5e, respectively. At $B = 0 \mathrm{T}$, the agreement with the WF law is good. However, as field increases to $B = 9 \mathrm{T}$, a four-fold violation of WF law is observed (Figure 5d). This happens across wide temperature range but not at low temperatures, indicating the link of scattering (Supplementary Information XI). The observed strong violation of the WF law hints at the possibility of field-driven, scattering-enhanced collective behaviors in a large entropic system, and is subject to further investigation.

## Discussion

**Pathway toward room-temperature topological thermoelectrics.** In this work, we report high thermopower and giant thermoelectric power factor in the WSM TaP, induced by the quantized thermoelectric Hall effect originating from topologically protected Weyl nodes at the quantum limit. These features are linked as follows: in a strong magnetic field, $S_{xx} \sim \alpha_{xy} \rho_{yx}$, the quantizing behavior of $\alpha_{xy}$ combined with the continual increase of $\rho_{yx}$ with magnetic field leads to the growth of $S_{xx}$, while the suppression of longitudinal portion $\alpha_{xx} \rho_{xx}$ to $S_{xx}$ further contributes to high power factor $S_{xx}^2 / \rho_{xx}$. The choice of TaP is due to its simpler Fermi surface compared to other members in the TaAs family[25]. On the other hand, the huge mass difference between Ta and P atoms reduces the three-phonon process and results in a high thermal conductivity, making it not directly applicable as a thermoelectric material. Even so, our work sheds light on a systematic pathway to seek promising topological thermoelectrics: To increase $S_{xx}$, large carrier compensation is desired (Eq. (1)). To simultaneously maintain low $\rho_{xx}$, simultaneously high carrier densities $n_h$ and $n_e$ are required but not sufficient. In a topologically trivial semimetal such as Bi, although high thermopower can be achieved at the quantum limit ( $S_{xx}(\text{Bi}) \sim 3 \times 10^3 \mu\text{VK}^{-1}$ vs $S_{xx}(\text{TaP}) \sim 1 \times 10^3 \mu\text{V/K}$ ), the electrical resistivity is significantly enhanced at high magnetic field ( $\rho_{xx}(\text{Bi}) \sim 2 \times 10^{-2} \Omega\text{m}$ vs $\rho_{xx}(\text{TaP}) \sim 1 \times 10^{-5} \Omega\text{m}$ )[10], indicating the crucial contribution of the gapless $n=0$ LL states from the topologically protected Weyl nodes. To tune the working temperature toward room temperature, long relaxation time is favored, along with preservation of the quantum limit, where thermal energy is smaller than the Landau level spacing, $k_B T << v_F \sqrt{\hbar e B}$ [21]. Finally, intrinsic magnetism can be used to replace the external $B$-field. Overall, we foresee that magnetic topological WSMs and related topological nodal line semimetals [32,33] with protected gapless states are promising candidate materials for thermoelectrics when charge carriers are largely compensated and the Fermi level is tuned to the gapless nodes to unlock the quantized thermoelectric Hall effect. To summarize, we demonstrated non-saturating longitudinal thermopower, giant power factor, and a signature of quantized thermoelectric Hall conductivity in a WSM in quantitative agreement with recent theoretical proposals. Furthermore, a field-driven breakdown of the WF law is observed at intermediate temperatures. Given the promising magnitudes of thermopower and power factor, our work sheds light on a few essential requirements that high-performance room-temperature thermoelectrics should meet. These include a way to create giant electronic entropy and reduce carrier density, and a way of evading localization while maintaining high electrical conductivity. Interestingly, the $n=0$ LL state with topologically-protected Weyl nodes in a WSM satisfies all these requirements. Our work thus demonstrates the possibility of topological materials to lead the breakthrough of thermoelectric materials working below room temperature.

**Note Added:** When we were finalizing this manuscript, we became aware of a work on Dirac semimetal[34]. The related work and our work mutually strengthened each other on the part of the quantized thermoelectric Hall effect.

## Acknowledgements

The authors thank S.Y. Xu for the helpful discussions. F. H., N.A, T.H. and M. L. thank the support from US DOE BES Award No. DE-SC0020148. N.A. acknowledges the support of the National Science Foundation Graduate Research Fellowship Program under Grant No. 1122374. T.N. thanks the support from the MIT SMA-2 Fellowship Program and Sow-Hsin Chen Fellowship. V.K., B.S. and L.F. thank support from DOE Office of Basic Energy Sciences, Division of Materials Sciences and Engineering under Award DE-SC0018945. Q.N. thanks the support from MIT NSE UROP Program. Z. Ding thanks support from DOD Defense Advanced Research Projects Agency (DARPA) Materials for Transduction (MATRIX) program under Grant HR0011-16-2-0041. R.P.P. thanks the support from FEMSA and ITESM. B.S. is supported by the NSF STC "Center for Integrated Quantum Materials" under Cooperative Agreement No. DMR-1231319. L.F. is partly supported by the David and Lucile Packard Foundation. This research on neutron scattering used resources at the High Flux Isotope Reactor, a DOE Office of Science User Facility operated by the Oak Ridge National Laboratory. This research used resources of the Advanced Photon Source, a U.S. Department of Energy (DOE) Office of Science User Facility operated for the DOE Office of Science by Argonne National Laboratory under Contract No. DE-AC02-06CH11357.

**Author contribution**

M.L. and F.H. designed the experiments. V.K., B.S. and L.F. formulated the theory. F.H. synthesized the samples and performed the transport measurements, with the help of T.N., Q.N and S.P. N.A, T.N. and F.H. analyzed the data, with contribution from all the authors. Z.D. carried out *ab initio* calculations. N.A., T.N., F.H., R.P.P., M.L along with A.A. E.A., S.C., J. F.-B. and S.H. performed inelastic scattering measurements. M.L., N.A., F.H., T.N., wrote the paper with input from all authors.

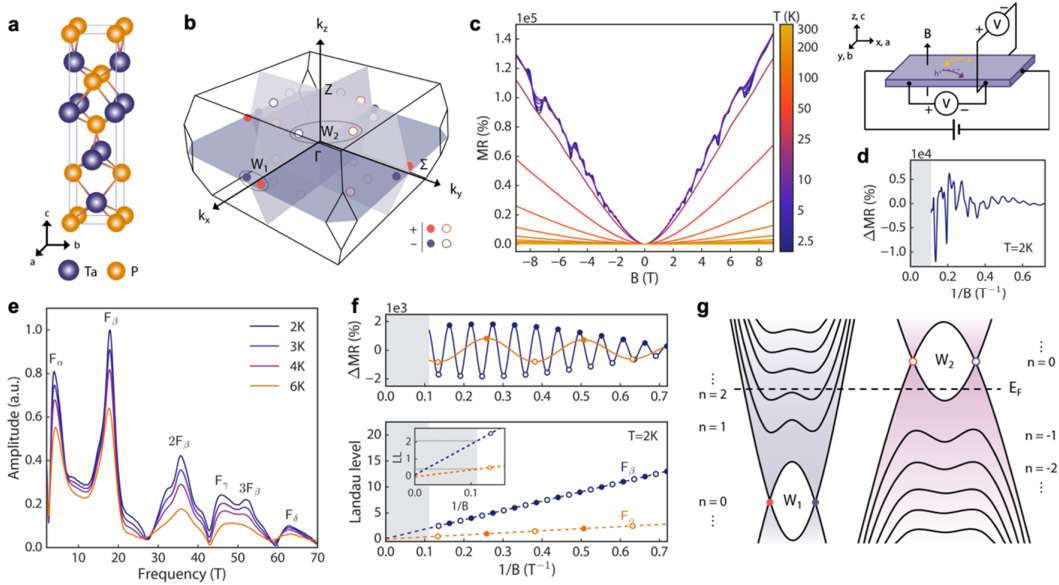

Figure 1. **Quantum oscillation of TaP. a** The inversion-symmetry-breaking crystal structure and **b** the Brillouin zone of TaP, highlighting the locations of the inequivalent Weyl nodes W1 (filled circles) and W2 (empty circles). The Weyl nodes are paired as source "+" (red) and sink "-" (blue) of Berry curvature, separated in momentum space. **c** Magnetoresistance (MR) as a function of magnetic field at different temperatures from 2.5K to 300K. A high (>$10^5$%) MR ratio is observed. **d** The MR measurement configuration (top) and $\Delta MR$ as a function of 1/B (bottom). e- and h+ denote electrons and holes, respectively. **e** The Fourier transform of $\Delta MR$ showing a low oscillation frequency $F_\alpha$ = 4T. This is a signature that, in addition to the electron pocket from W1 Weyl node contributing to $F_\beta$ = 18T, we are very close to the W2 Weyl node. **f** The SdH oscillation and Landau level index plot, from which we obtained an $n$=2 Landau level and another $n=0$ Landau level. **g** The schematic bandstructure at finite magnetic fields of our TaP sample.

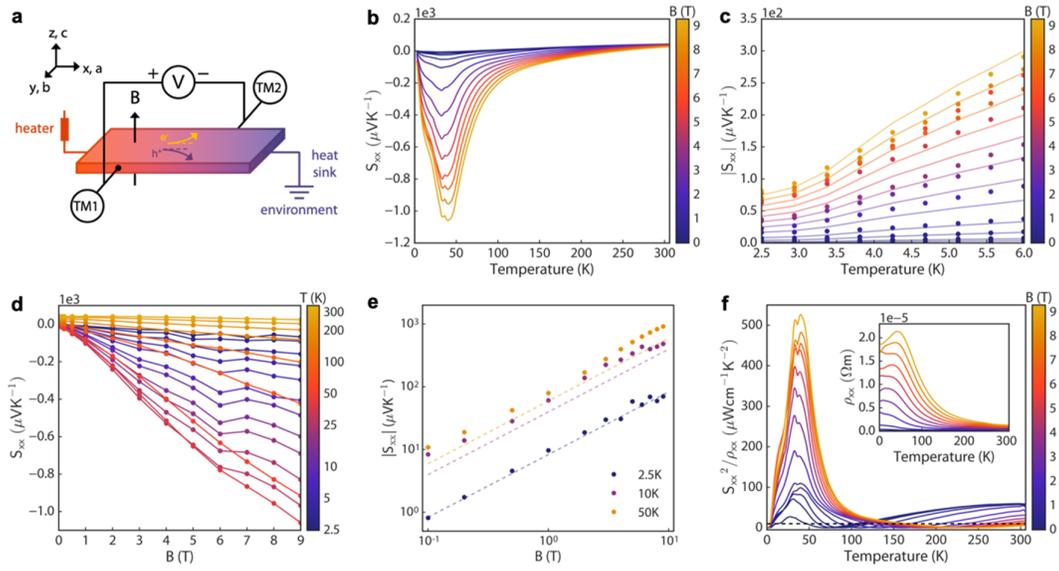

Figure 2. **Non-saturating thermopower at high fields. a** The schematics of the diagonal offset thermoelectric measurement geometry. TM1 and TM2 represent thermometer 1 and 2. The temperature difference between the short ends of the sample is represented by the color gradient from red (high) to blue (low). **b** Longitudinal thermopower $S_{xx}$ as a function of temperature at various fields. The double peaks emerge at ~33K and ~40K. **c** $S_{xx}$ in the low-temperature range, showing the quasi-linear growth as a function of temperature. **d** $S_{xx}$ replotted as a function of $B$, showing unbounded linear growth with field. The onset of the linear behavior indicates entrance into the quantum limit regime. The oscillatory behavior ~20K at $B$=6T is caused by the quantum oscillation effect. **e** $S_{xx}$ as a function of $B$ at a few representative temperatures. The dashed lines are theoretical values using Eq. (1) by substituting the fitted $v_F$ from Eq. (3) (for T=2.5K) and Eq. (S14) (for T≥10K). **f** The power factor as a function of temperature. The black-dashed line is a reference peak value for SnSe.

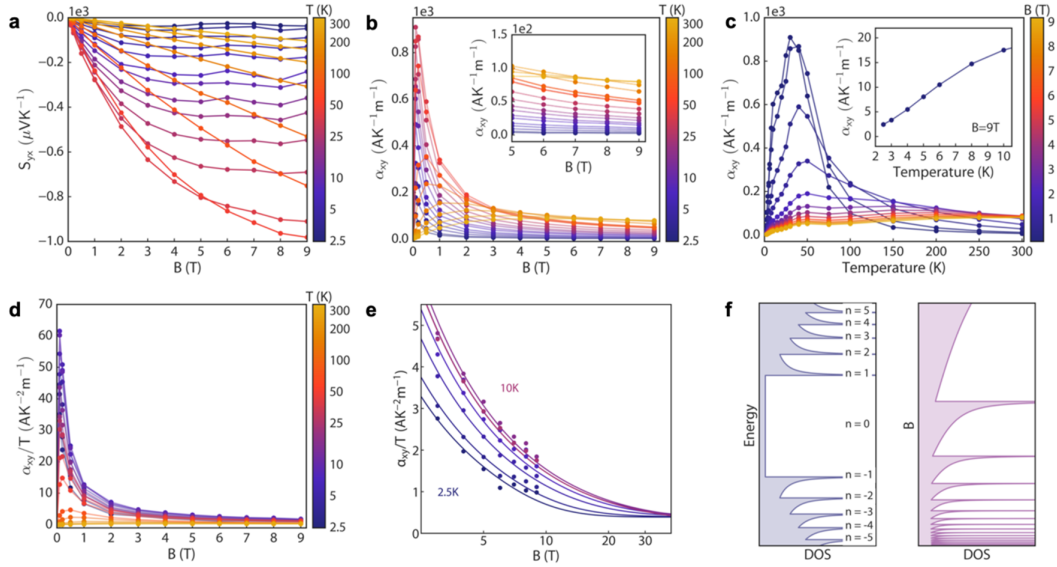

Figure 3. **The Quantized thermoelectric Hall effect. a** Transverse thermopower $S_{yx}$ as a function of magnetic field at different temperatures. **b** Thermoelectric Hall conductivity $\alpha_{xy}$ as a function of magnetic field at different temperatures. The peak value is caused by the finite scattering effect. **c** Thermoelectric Hall conductivity $\alpha_{xy}$ as a function of temperature at various fields. The inset shows a linear behavior of $\alpha_{xy}$ versus $T$ at low temperatures. **d** $\alpha_{xy}/T$ as a function of magnetic field at various temperatures. **e** An extrapolation of **d** showing a convergence to the quantized value at low temperatures. **f** The density of states (DOS) of each Landau level (LL), highlighting the unique $n=0$ LL in a WSM. At high-enough $B$, $n=0$ LL drives the DOS $\propto B$.

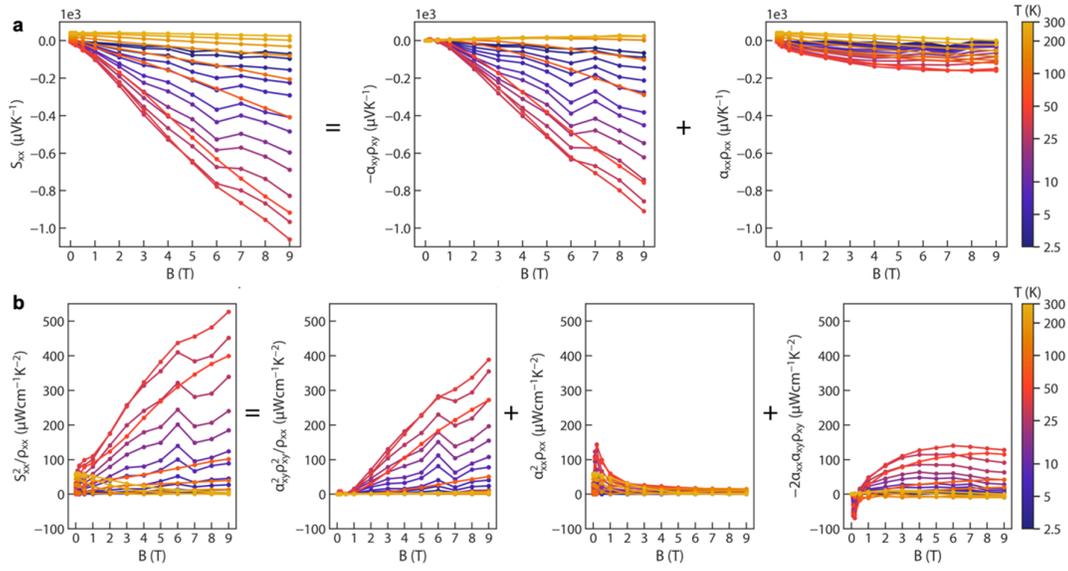

Figure 4. **Dominant contribution from the transverse thermoelectric Hall conductivity to longitudinal thermopower and power factor**. **a** Total $S_{xx}$ (left hand side, LHS) at various temperatures as a function of magnetic field, and its transverse component $-\alpha_{xy}\rho_{xy}$ (1st term on the right-hand-side, RHS) and longitudinal contribution $+\alpha_{xx}\rho_{xx}$ (2nd term on the RHS). **b** Total $S_{xx}^2/\rho_{xx}$ (LHS) at various temperatures as a function of magnetic field, and its transverse component $+\alpha_{xy}^2\rho_{xy}^2/\rho_{xx}$ (1st term on RHS), longitudinal contribution $+\alpha_{xx}^2\rho_{xx}$ (2nd term on the RHS), and the cross term $-2\alpha_{xy}\alpha_{xx}\rho_{xy}$ (3rd term on the RHS). The dominant contribution of transverse component can be seen at all temperatures.

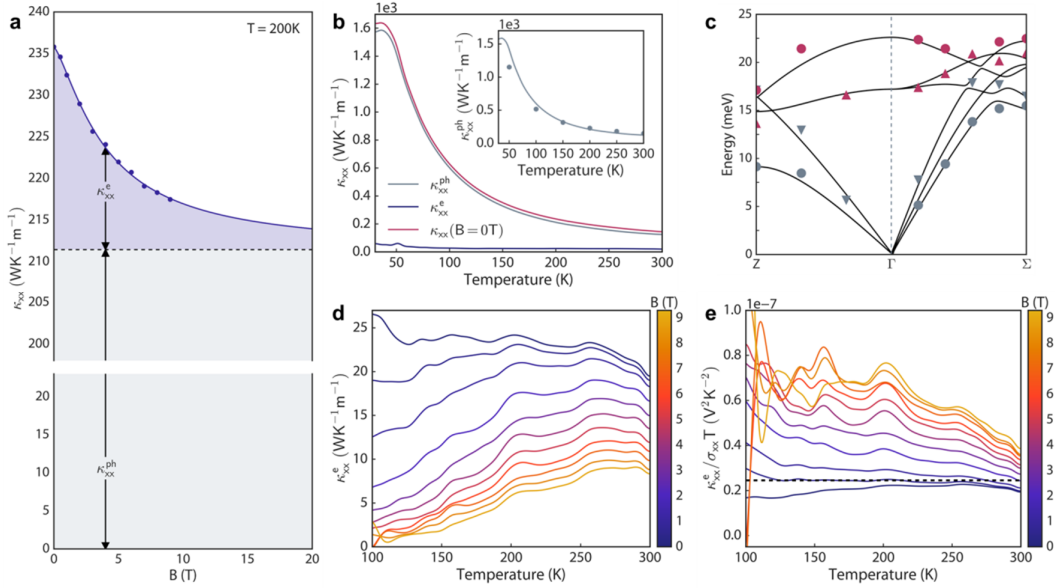

**Figure 5. The Wiedemann-Franz Law. a** The schematics of the separation process of electronic thermal conductivity $\kappa_{xx}^{e}$ from the lattice thermal conductivity $\kappa_{xx}^{ph}$ using the field-dependence. **b** Separation of phonon and electronic contributions to the longitudinal thermal conductivity with inset displaying a computation (scattered points) of the phonon thermal conductivity from first principles. **c** Experimentally measured values of phonon modes (scattered points) of TaP along high-symmetry line Z-Γ-Σ taken by inelastic x-ray and neutron scattering with accompanying *ab initio* calculation (solid lines), displaying good agreement between *ab initio* calculations and experiment. **d** The electronic contribution of the thermal conductivity as a function of temperature at various fields. **e** The Lorenz number as a function of temperature at various fields. The black line indicates the theoretical value of the Wiedemann-Franz law.

# Quantized Thermoelectric Hall Effect Induces Giant Power Factor in a Topological Semimetal: Supplementary Information

Fei Han[1*†], Nina Andrejevic[2†], Thanh Nguyen[1†], Vladyslav Kozii[3†], Quynh T. Nguyen[1,3], Tom Hogan[4], Zhiwei Ding[2], Ricardo Pablo-Pedro[1], Shreya Parjan[5], Brian Skinner[3], Ahmet Alatas[6], Ercan Alp[6], Songxue Chi[7], Jaime Fernandez-Baca[7], Shengxi Huang[8], Liang Fu[3*], Mingda Li[1*]

[1]Department of Nuclear Science and Engineering, Massachusetts Institute of Technology, Cambridge, MA 02139, USA

[2]Department of Materials Science and Engineering, Massachusetts Institute of Technology, Cambridge, MA 02139, USA

[3]Department of Physics, Massachusetts Institute of Technology, Cambridge, MA 02139, USA

[4]Quantum Design Inc, San Diego, CA 92121

[5]Department of Physics, Wellesley College, 106 Central St, Wellesley, MA 02481, USA

[6]Advanced Photon Source, Argonne National Laboratory, Lemont, IL 60439, USA

[7]Neutron Scattering Division, Oak Ridge National Laboratory, Oak Ridge, TN, 37831, USA

[8]Department of Electrical Engineering, The Pennsylvania State University, State College, PA 16802, USA

†These authors contribute equally to this work.

*Corresponding authors: hanfei@mit.edu; liangfu@mit.edu; mingda@mit.edu.

**Contents**

## I.  High-quality Single-crystal Growth

We successfully obtained centimeter-sized single crystals of TaP using the vapor transport method. The single crystals of TaP were prepared in two steps. In the first step, 3 grams of Ta (Beantown Chemical, 99.95%) and P (Beantown Chemical, 99.999%) powders were weighed, mixed, and ground in a glovebox. The mixed powders were flame-sealed in a quartz tube which was subsequently heated to 700ºC and dwelled for 20 hours for a pre-reaction. In the second step, the obtained TaP powders were sealed in another quartz tube with 0.4 grams of $I_2$ (Sigma Aldrich, >=99.8%) added. The tube containing TaP and $I_2$ was then horizontally placed in a two-zone furnace. To improve the crystal size and quality, instead of setting a 100ºC temperature difference, we gradually increased the temperature difference from zero until the $I_2$ transport agent started to flow. This process seems to be furnace- and distance-specific. In our case, the optimal temperatures for the two zones are 900ºC and 950ºC, respectively, and the distance between the two heating zones is constantly optimized. With the help of the transport agent $I_2$, the TaP source materials transferred from the cold end of the tube to the hot end and condensed at the hot end in a single-crystalline form in 14 days. The resulting products of TaP single crystals are centimeter-sized and have a metallic luster. A typical crystal is shown in Figure S1.

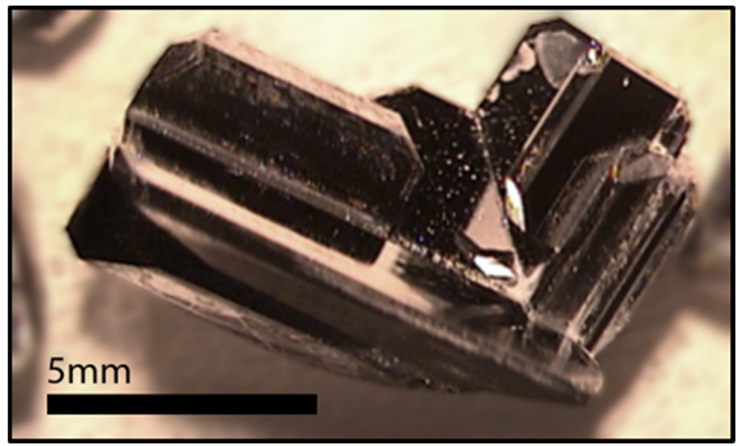

Figure S1. **Single crystals of TaP grown by the vapor transport method.**

## II.     Sample Preparation for Measurements

To conduct high-precision electrical and thermal transport measurements on TaP, we performed a thinning-down process on the crystals. Due to the very high electrical and thermal conductivities of TaP, it is difficult to do high-precision electrical and thermal transport measurements on the as-grown crystals. To magnify the electrical resistance and the temperature gradient in the electrical and thermal transport measurements, one piece of crystal was polished down to a thin slab along the *c*-axis. Figures S2a and b display top and side views of the thinned-down crystal we used for the thermal transport measurement (namely thermoelectric measurement) whose thickness is only 0.17 mm. Figure S2c shows the probe configuration on the thinned-down crystal for the thermoelectric measurement, and Figures S2d and e give explanatory schematics to the usages of the probes in the thermal conductivity, thermopower, and resistivity acquisitions. The contacts of the probes were made with the silver epoxy H20E.

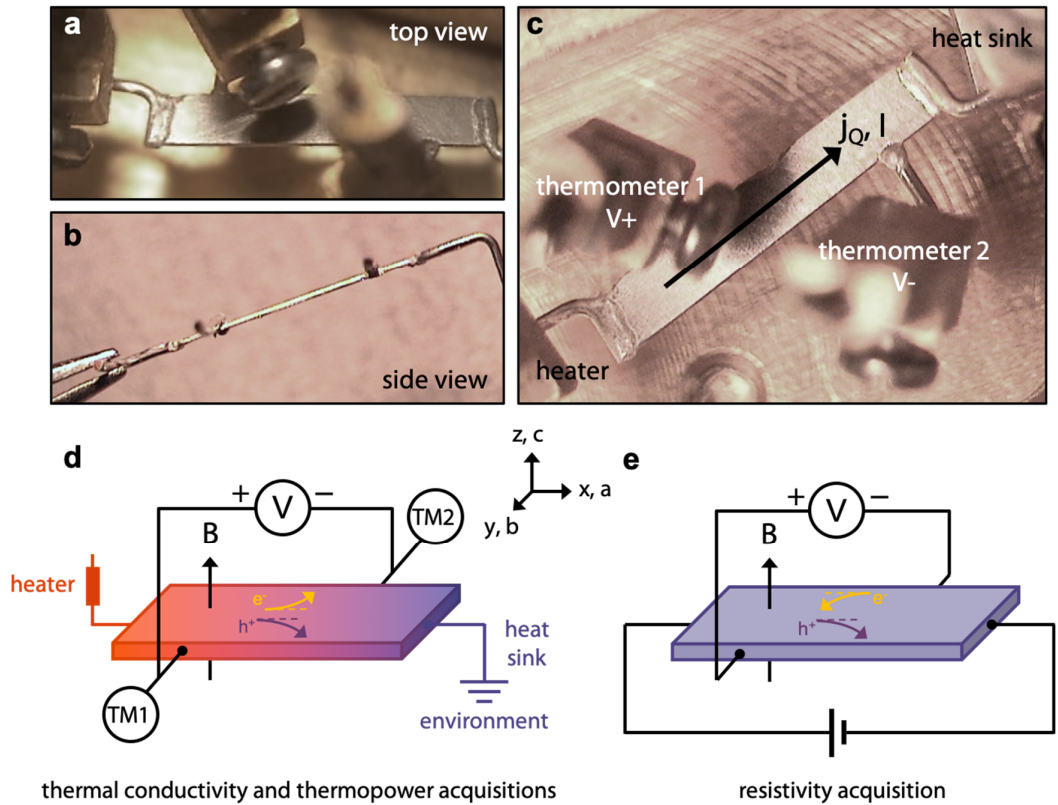

Figure S2. **a** Top and **b** side view of the thinned-down TaP crystal we used for the thermoelectric measurement. The thickness is as thin as 0.17mm. **c** Probe configuration on the thinned-down crystal for the thermoelectric measurements. Explanatory schematic diagrams for the usage of the probes **d** in the thermal conductivity and thermopower acquisitions and **e** in the resistivity acquisition. TM1 and TM2 represent thermometer 1 and 2. The temperature difference between the short ends of the sample in **d** is represented by the color gradient from red (high) to blue (low). e- and h+ denote electrons and holes, respectively.

## III.  Carrier Concentration and Mobility

The electrical and thermal transport measurements were carried out with the electrical transport option (ETO) and the thermal transport option (TTO) of physical property measurement system (PPMS), respectively. The data about the quantum oscillations were measured with the ETO whereas the data about the thermoelectric (including resistivity) with the TTO. When we performed the ETO measurements we adopted a standard six-probe configuration and connected the longitudinal and transverse probes to two independent measurement channels. The details about the ETO measurement can be found in Figure S3a. However, because the TTO has only one measurement channel, to measure the longitudinal and transverse thermal conductivities ($\kappa_{xx}$ and $\kappa_{yx}$), resistivities ($\rho_{xx}$ and $\rho_{yx}$), and Seebeck coefficients ($S_{xx}$ and $S_{yx}$) simultaneously, we used a diagonal offset probe geometry for the thermal transport measurement, as shown in Figures S7a and S8a. For the detailed description about the TTO measurement, consult Supplementary Information VI.

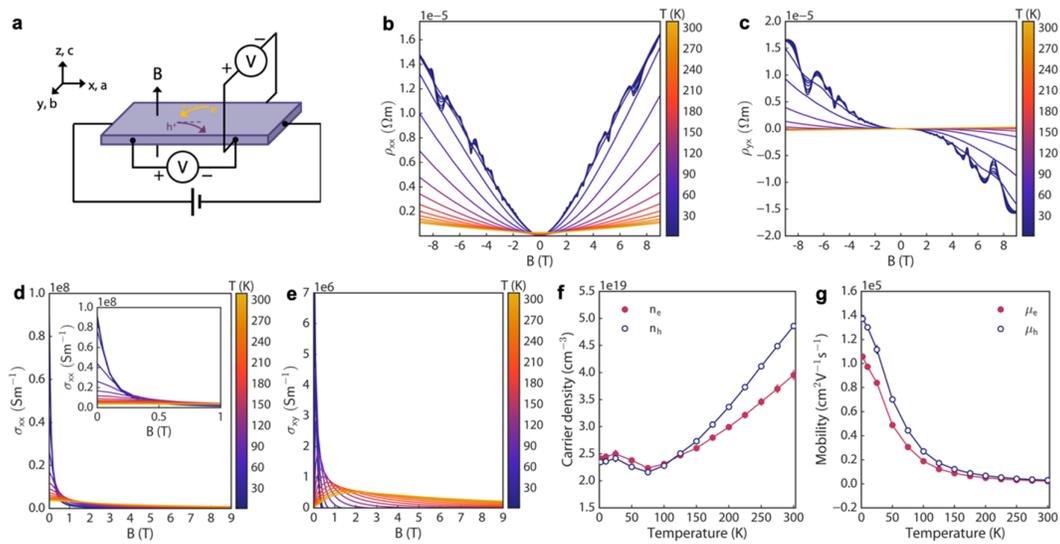

Figure S3. **a** Schematic diagram of the electrical transport measurement in the six-probe geometry. e- and h+ denote electrons and holes, respectively. Longitudinal and transverse resistivities and conductivities **b** $\rho_{xx}$, **c** $\rho_{yx}$, **d** $\sigma_{xx}$, and **e** $\sigma_{xy}$ as functions of magnetic field at different temperatures. **f** Carrier concentration and **g** mobility of electrons and holes resulting from the two-band model fitting.

To experimentally validate the prediction of a quantized thermoelectric Hall effect requires information about the carrier concentration and mobility. To extract this information, we carried out a delicate electrical transport measurement with the ETO of the PPMS. The measurement was done using a standard six-probe geometry, schematically shown in Figure S3a. With the symmetric probe configuration, the measured longitudinal resistivity $\rho_{xx}$ is symmetric with respect to the applied magnetic field, while the transverse resistivity $\rho_{yx}$ is antisymmetric, as shown in Figure S3b and c. In both $\rho_{xx}$ and $\rho_{yx}$, strong Shubnikov-de Haas (SdH) oscillations can be observed at low temperatures. The oscillation is preserved up to 25K, indicating high-quality crystallization in our sample, as the temperature damping effect would otherwise eliminate the quantum oscillation at this relatively high temperature.

Because the contacts on the sample were made manually with silver epoxy, the measured data exhibit slight asymmetry due to slight misalignment of the contacts. To eliminate the effect of the contact misalignment, we averaged the $\rho_{xx}$ and $\rho_{yx}$ using the equations listed below:

$$\rho_{xx}(B) = \frac{\rho_{xx}(+B) + \rho_{xx}(-B)}{2}, \quad \rho_{yx}(B) = \frac{\rho_{yx}(+B) - \rho_{yx}(-B)}{2}. \tag{S1}$$

Then we calculated the longitudinal and transverse conductivities $\sigma_{xx}$ and $\sigma_{xy}$ using the following equations:

$$\sigma_{xx} = \frac{\rho_{xx}}{\rho_{xx}^2 + \rho_{xy}^2}, \quad \sigma_{xy} = -\frac{\rho_{xy}}{\rho_{xx}^2 + \rho_{xy}^2} = \frac{\rho_{yx}}{\rho_{xx}^2 + \rho_{yx}^2}. \tag{S2}$$

The field dependence of $\sigma_{xx}$ and $\sigma_{xy}$ at various temperatures is shown in Figures S3d and e. To extract the carrier concentration and mobility, we simultaneously fit the $\sigma_{xx}$ and $\sigma_{xy}$ data as functions of $B$ using a two-band model defined by:

$$\sigma_{xx} = \frac{n_e \mu_e e}{1+(\mu_e B)^2} + \frac{n_h \mu_h e}{1+(\mu_h B)^2}$$
$$\sigma_{xy} = \left[ n_h \mu_h^2 \frac{1}{1+(\mu_h B)^2} - n_e \mu_e^2 \frac{1}{1+(\mu_e B)^2} \right] eB \qquad (S3)$$

where $n_e$ and $n_h$ denote the electron and hole carrier densities, $\mu_e$ and $\mu_h$ are the corresponding mobilities, and $e$ is the elementary charge. We thereby extract the electron and hole carrier densities and mobilities as functions of temperature, as shown in Figures S3f and g. The electron and hole concentrations are nearly compensated at low temperatures. This proves the origin of the giant magnetoresistance.

## IV. Analysis of Quantum Oscillation

Since the carrier pockets analysis based on quantum oscillation can be influenced by the choice of background of magnetoresistance (MR), we adopted three independent methods using 1) background-free curvature approach (Figure S4), 2) a $T$=25K data without quantum oscillation as background (Figure S5), and 3) a fitted background to a linear-quadratic function (Figure S6), all of which lead to a consistent conclusion of the existence of a low frequency carrier pocket $F_\alpha = 2.3T \sim 4T$. This enables the possibility that the carrier pockets of W2 Weyl point can indeed reach the desired $n$=0 LL.

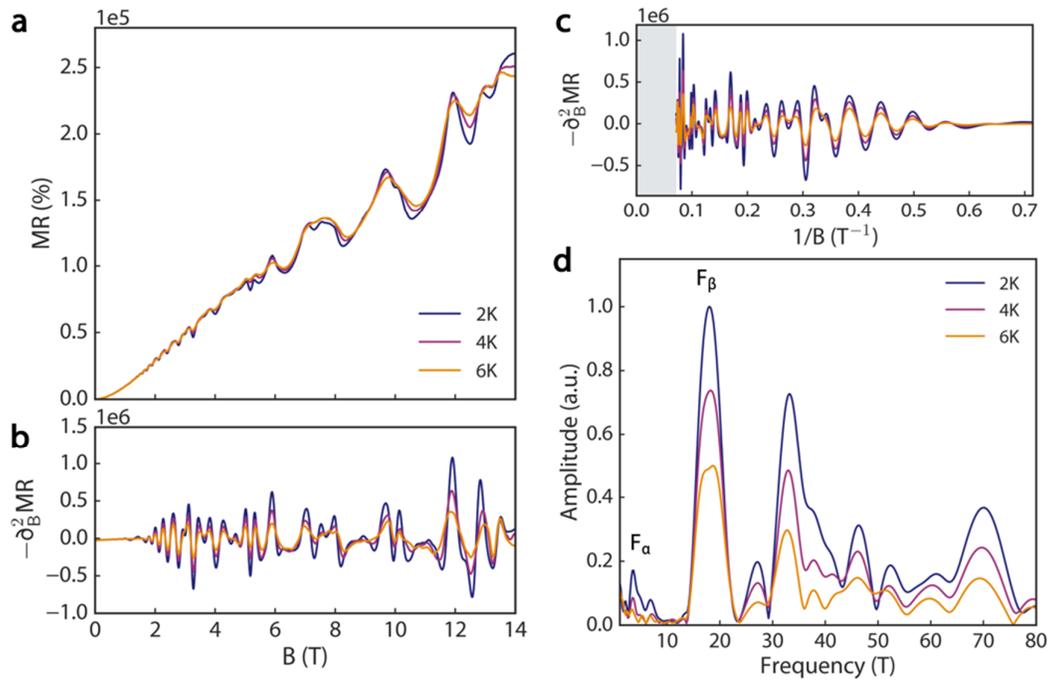

Figure S4. **MR analysis using curvature-based background subtraction.** In this approach, a second-order derivative against magnetic field $B$ is taken to the MR data, where all linear, constant, and quadratic terms will be automatically wiped out without need to manually choosing background. Although this method is seldom used, this may offer an alternative but strong approach for MR analysis. **a** MR data up to $B$=14T, at $T$=2K, 4K and 6K. $\partial_B^2 MR$ as a function of **b** $B$ and **c** $1/B$. **d** Fourier transform of (c), showing the two carrier pockets $F_\alpha = 3.8T$ and $F_\beta = 18T$.

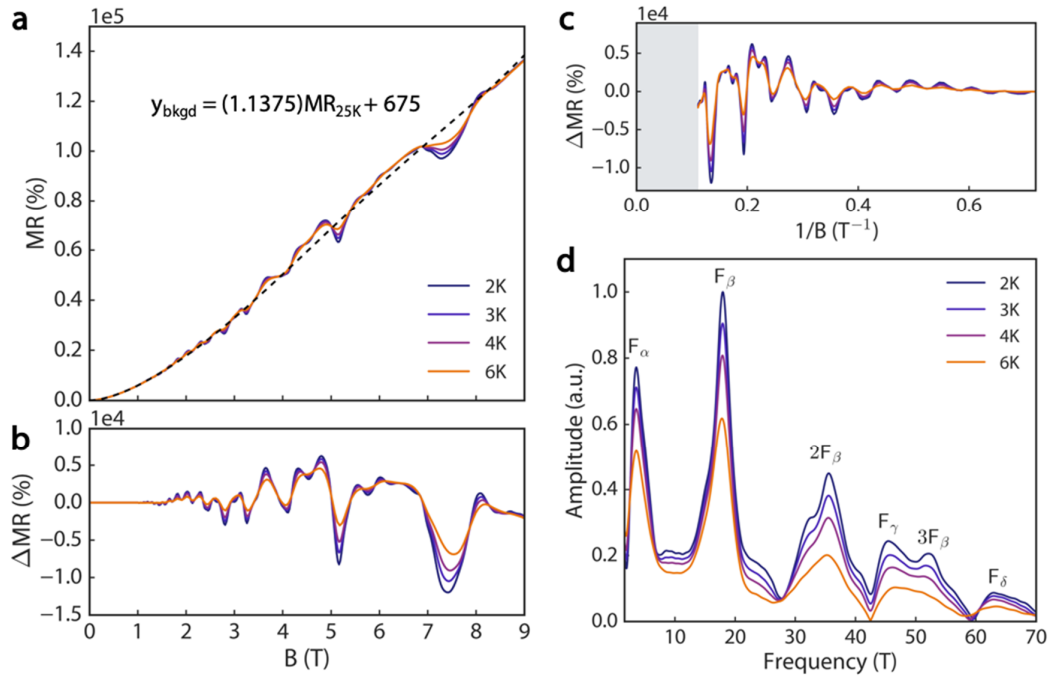

Figure S5. **MR analysis using *T*=25K data as background.** In this approach, since the quantum oscillation fully disappears at *T*=25K, we can use the MR at *T*=25K as the background. **a** The low-temperature MR data and the background using linearly transformed MR at 25K (black dashed line). The $\Delta MR$ after background subtraction in terms of **b** *B* and **c** *1/B*. **d** The Fourier transform of (c), showing the two carrier pockets $F_\alpha = 4T$ and $F_\beta = 18T$.

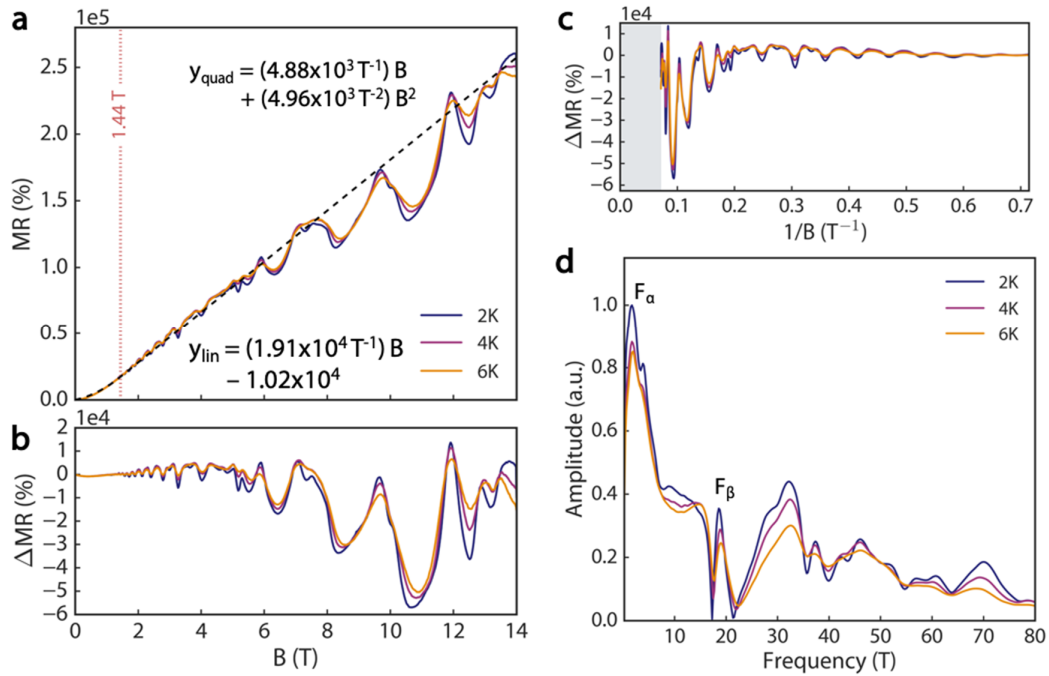

Figure S6. **MR analysis using polynomial fitting-based background subtraction.** In this approach, an optimal threshold $B_0$ is obtained by curve-fitting, which divides the measurement into a quadratic ( $B < B_0$ ) and linear ( $B > B_0$ ) regime, with their magnitudes and slopes matched at $B_0$. **a** The low-temperature MR data and a fitted linear-quadratic function as background (black dashed line). The red vertical dashed line indicates the fitted value $B_0 = 1.44$T. The $\Delta MR$ after background subtraction in terms of **b** $B$ and **c** $1/B$. **d** The Fourier transform of **c**, showing the two carrier pockets $F_\alpha = 2.3T$ and $F_\beta = 19T$. Despite a small quantitative difference, the existence of the low-frequency carrier pocket is confirmed.

## V. Landau Level and Quantum Limit

The quantized thermoelectric Hall effect considered in this work is theoretically predicted to exist in the quantum limit of Dirac/Weyl semimetals[S1,2]. Therefore, to examine the validity of the theoretical prediction, we first verify that the quantum limit condition is satisfied by Weyl fermions in our TaP sample. To do this, we performed a thorough analysis of the quantum oscillations observed in the electrical transport measurement, as shown in Figure 1 in the main text and discussed in detail in the previous section. The quantum oscillation data ΔMR shown in Figure 1d of the main text was obtained by subtracting a smooth background from the magnetoresistance (MR) data, Figure 1c, where MR is defined according to:

$$MR = \frac{\rho_{xx}(B) - \rho_{xx}(B=0\text{T})}{\rho_{xx}(B=0\text{T})} \times 100\%.$$

(S4)

From the fast Fourier transform (FFT) analysis depicted in Figure 1e, we observe four noticeable oscillation frequencies: $F_\alpha = 4\text{T}$, $F_\beta = 18\text{T}$, $F_\gamma = 46\text{T}$, and $F_\delta = 64\text{T}$. After performing a standard signal filtering process by performing inverse FFT to the two relatively low frequencies of 4T and 18T individually, we isolate the two oscillation components from the pristine data and determine the corresponding Landau levels (LLs) by assigning an integer (half-integer) value to the oscillation maxima (minima), as shown in Figure 1f. From the LL index fan, we conclude that in our TaP sample, the $\alpha$ Fermi pocket corresponding to the 4T frequency is in the $n=0$ LL at our maximum field of $B=9\text{T}$, whereas the $\beta$ Fermi pocket corresponding to the 18T frequency is in the $n=2$ LL. Specifically, the $\alpha$ Fermi pocket enters the quantum limit (lowest LL) approximately at 3.8T, and the $\beta$ Fermi pocket will reach the quantum limit at an approximate field of 16T. The linear fitting of the LL index as a function of 1/B yields intercepts of -0.037 and 0.065 for $\alpha$ and $\beta$, respectively. Both are in the range of -1/8 to 1/8, proving the bands in the $\alpha$ and $\beta$ Fermi pockets are topologically non-trivial and thus Weyl cones are present[S3]. From this, we can further conclude that the Weyl fermions in the smallest Fermi pocket of TaP are well within the quantum limit at our maximum applied field, whereas the Weyl fermions in the second smallest Fermi pocket are nearing the onset of the quantum limit.

## VI. Data Analysis for Thermoelectric Measurement

Figure S7a schematically shows the principle behind the thermal transport measurement in the diagonal offset probe geometry. Using the TTO of the PPMS, the heater on the left end of the thinned-down crystal and heat sink on the right establish a continuous heat flow along the $a$ or $b$ axis ($a$ and $b$ are equivalent for this tetragonal system), as shown in Figure S7. The thermal conductivity is directly calculated by the PPMS using the applied heater power, the resulting temperature difference $\Delta T$ detected between the two thermometers, and the sample dimension. The voltage drop $\Delta V$ between the two thermometers is monitored simultaneously, which yields the Seebeck signals by calculation of $-\Delta V / \Delta T$. A magnetic field was applied along the $c$ axis for detecting the proposed quantized thermoelectric Hall effect. Figure S7b shows the temperature dependence of thermal conductivity of TaP at 9T and -9T. From this plot, we note that the thermal conductivities at positive and negative magnetic fields have a very slight difference. This indicates that the thermoelectric Hall effect (the transverse movement of thermal electrons in the presence of a magnetic field) provides a negligible but observable heat flow along the transverse direction. To extract the longitudinal thermal conductivity from the measured thermal conductivity, we use the following equations:

$$\rho_{th,xx}(B) = \frac{\kappa_{meas}(+B) + \kappa_{meas}(-B)}{2\kappa_{meas}(+B)_{meas}\kappa(-B)}, \quad \rho_{th,yx}(B) = \frac{\kappa_{meas}(+B) - \kappa_{meas}(-B)}{2\kappa_{meas}(+B)\kappa_{meas}(-B)} \times \frac{L}{W}, \quad (S5)$$

and

$$\kappa_{xx} = \frac{\rho_{th,xx}}{\rho_{th,xx}^2 + \rho_{th,yx}^2}, \quad (S6)$$

where $L$ and $W$ represent the length-wise and the width-wise separation between the two thermometers. Figure S7c displays the obtained longitudinal thermal conductivity $\kappa_{xx}$ as a function of temperature at different magnetic fields. From the inset of Figure S7c, we see that the applied magnetic field gradually suppresses the longitudinal thermal conductivity. This phenomenon is consistent with the giant magnetoresistance, as both originate from the greatly elevated electron scattering induced by the magnetic field. The magnitude of the thermal conductivity of TaP is very large compared to most materials, which explains the importance of thinning the sample prior to measurement. The Seebeck signals at 0T, 9T and -9T are plotted in Figure S7d, from which giant magnetic field-induced Seebeck signals can be observed at 9T and -9T. The data for 9T and -9T are asymmetrical due to the mutual presence of longitudinal and transverse Seebeck signals. We use the following equations to calculate the longitudinal and transverse Seebeck coefficients $S_{xx}$ and $S_{yx}$:

$$S_{xx}(B) = \frac{S_{meas}(+B) + S_{meas}(-B)}{2}, \quad S_{yx}(B) = \frac{S_{meas}(+B) - S_{meas}(-B)}{2} \times \frac{L}{W}. \quad (S7)$$

The temperature dependence of $S_{xx}$ and $S_{yx}$ collected at different magnetic fields is presented in Figures S7e and f. It is obvious that the applied magnetic fields induce giant Seebeck coefficients along both longitudinal and transverse directions. The longitudinal Seebeck coefficient $S_{xx}$ does not appear to saturate with increasing field up to the highest measured field of 9T. By contrast, $S_{yx}$ tends to saturate at high magnetic fields. Another novel behavior in $S_{xx}$ and $S_{yx}$ is the presence of a double-peak feature around $T$=40K. We provide a clear explanation of this feature in the main text.

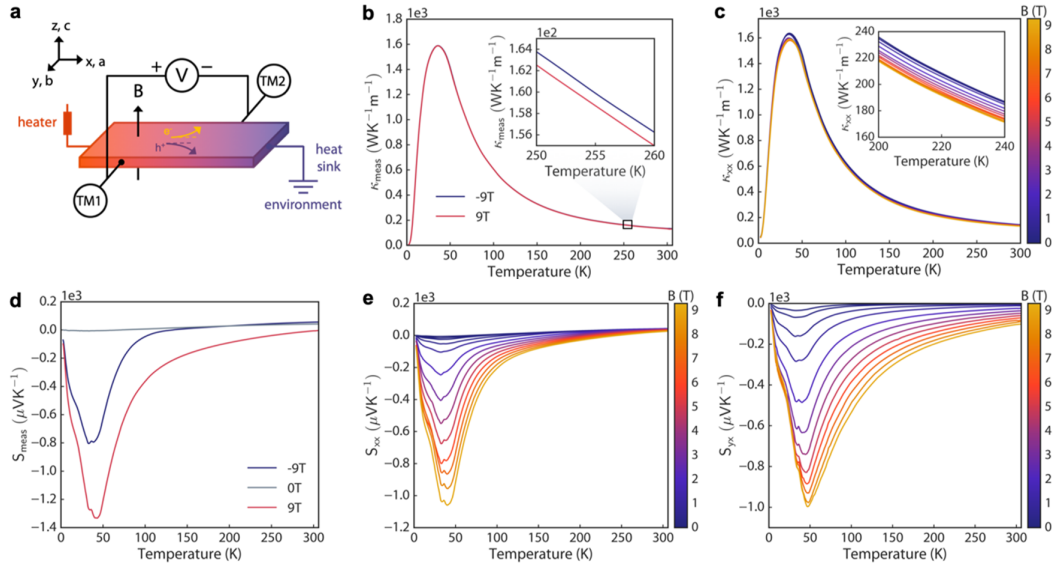

Figure S7. **a** Schematic diagram of the thermal transport measurement. TM1 and TM2 represent thermometer 1 and 2. The temperature difference between the short ends of the sample is represented by the color gradient from red (high) to blue (low). e- and h+ denote electrons and holes, respectively. **b** Thermal conductivities of TaP at 9T and -9T. **c** Longitudinal thermal conductivity of TaP as a function of temperature at various fields. **d** Measured Seebeck signals at 0T, 9T and -9T for the diagonal offset probe geometry. **e** Longitudinal and **f** transverse Seebeck coefficients $S_{xx}$ and $S_{yx}$ as functions of temperature at different magnetic fields.

After performing the thermal transport measurement at a certain temperature, a subsequent electrical transport measurement at the same temperature is made with the TTO. The inset of Figure S8a shows the schematic diagram for the electrical transport measurement in the diagonal offset geometry. In the presence of a magnetic field, the system applies an electrical current along the $a$ or $b$ axis, and the voltmeter between the diagonal offset probes detects the voltage drop which contains both longitudinal and transverse components. The longitudinal resistivity $\rho_{xx}$ and the transverse resistivity (also called Hall resistivity) $\rho_{yx}$ are separated using the following equations:

$$\rho_{xx}(B) = \frac{\rho_{\text{meas}}(+B) + \rho_{\text{meas}}(-B)}{2}, \quad \rho_{yx}(B) = \frac{\rho_{\text{meas}}(+B) - \rho_{\text{meas}}(-B)}{2} \times \frac{L}{W}. \quad (S8)$$

Figure S8a displays the measured resistivity at 0T, 9T and -9T. The disagreement between the 9T and -9T data is evidence of the mutual presence of the longitudinal and transverse resistivities $\rho_{xx}$ and $\rho_{yx}$. After separating $\rho_{xx}$ and $\rho_{yx}$ using Eq. (S8), as shown in Figures S8b and c, we then calculated the figure of merit $zT$ according to:

$$zT = \frac{S_{xx}^2 T}{\rho_{xx}\kappa_{xx}}. \quad (S9)$$

From the plot of $zT$ in Figure S8d, we note that, although the power factor (shown in Figure 2f in the main text) is record-breaking in magnitude, the $zT$ does not attain a very high value due to the significant thermal conductivity.

It should be noted that the giant magnetic field-induced Seebeck coefficients cannot be observed in the case of $B\|a\|j_Q$, which is evidenced by comparison of two geometries in Figures S8e and f. This indicates that the giant magnetic field-induced longitudinal and

transverse Seebeck coefficients in the case of $B\|c\perp j_Q$ originate from the quantized protection of the thermoelectric Hall Effect.

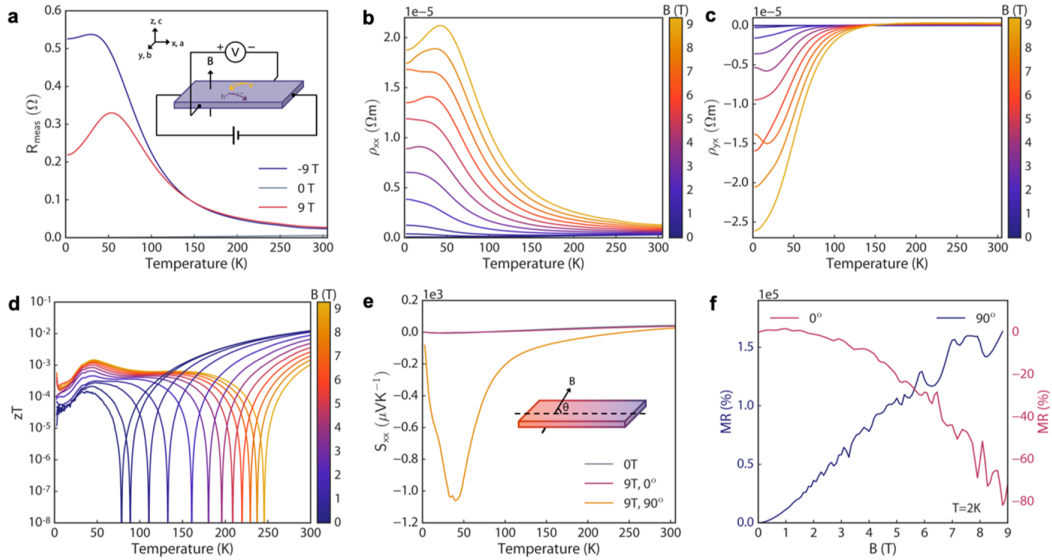

Figure S8. **a** Measured resistivities of TaP at 0T, 9T and -9T for the diagonal offset probe geometry. Inset: Schematic diagram of the electrical transport measurement. Longitudinal and transverse resistivities **b** $\rho_{xx}$, **c** $\rho_{yx}$, and **d** zT as functions of temperature at different magnetic fields. Comparison of the $B\|a\|j_Q$ and $B\|c\perp j_Q$ geometries for **e** $S_{xx}$ and **f** *MR*. The giant Seebeck coefficients were not observed in the $B\|a\|j_Q$ case.

To summarize the phase relations of various thermoelectric quantities, we plot the resistivity, thermopower, thermal conductivity, and thermoelectric Hall conductivity in both longitudinal and transverse directions and highlight their phase relations, as done in Figures S9-S13.

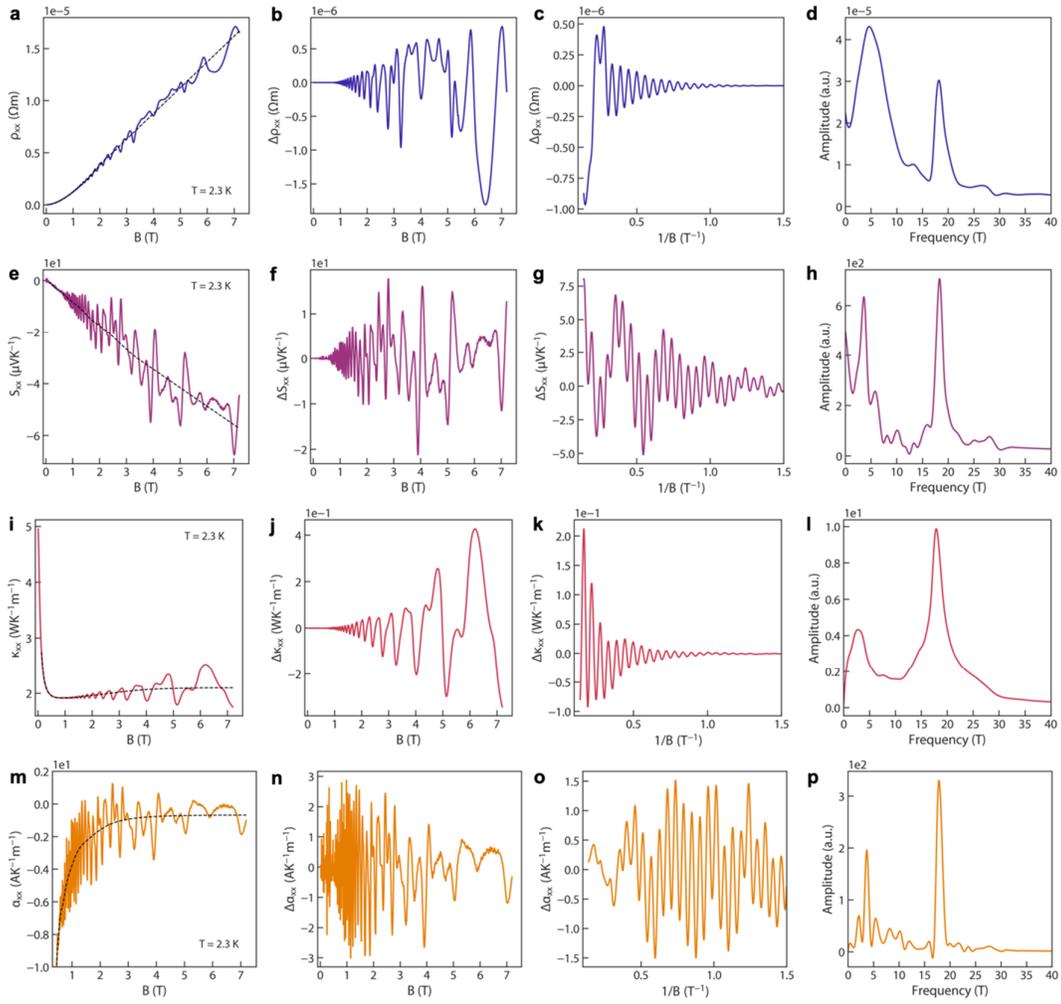

Figure S9. **Quantum oscillation of longitudinal thermoelectric properties at base temperature. a-d** Longitudinal resistivity $\rho_{xx}$. **a** $\rho_{xx}$ vs $B$, **b** the background-subtracted part $\Delta\rho_{xx}$ vs $B$, **c** $\Delta\rho_{xx}$ vs $1/B$, and **d** the Fourier transform of **c**. **e-h** Longitudinal thermopower $S_{xx}$. **e** $S_{xx}$ vs $B$, **f** the background-subtracted part $\Delta S_{xx}$ vs $B$, **g** $\Delta S_{xx}$ vs $1/B$, and **h** the Fourier transform of **g**. **i-l** Longitudinal thermal conductivity $\kappa_{xx}$. **i** $\kappa_{xx}$ vs $B$, **j** the background-subtracted part $\Delta\kappa_{xx}$ vs $B$, **k** $\Delta\kappa_{xx}$ vs $1/B$, and **l** the Fourier transform of **k**. **m-p** Longitudinal thermoelectric conductivity $\alpha_{xx}$. **m** $\alpha_{xx}$ vs $B$, **n** the background-subtracted part $\Delta\alpha_{xx}$ vs $B$, **o** $\Delta\alpha_{xx}$ vs $1/B$, and **p** the Fourier transform of **o**.

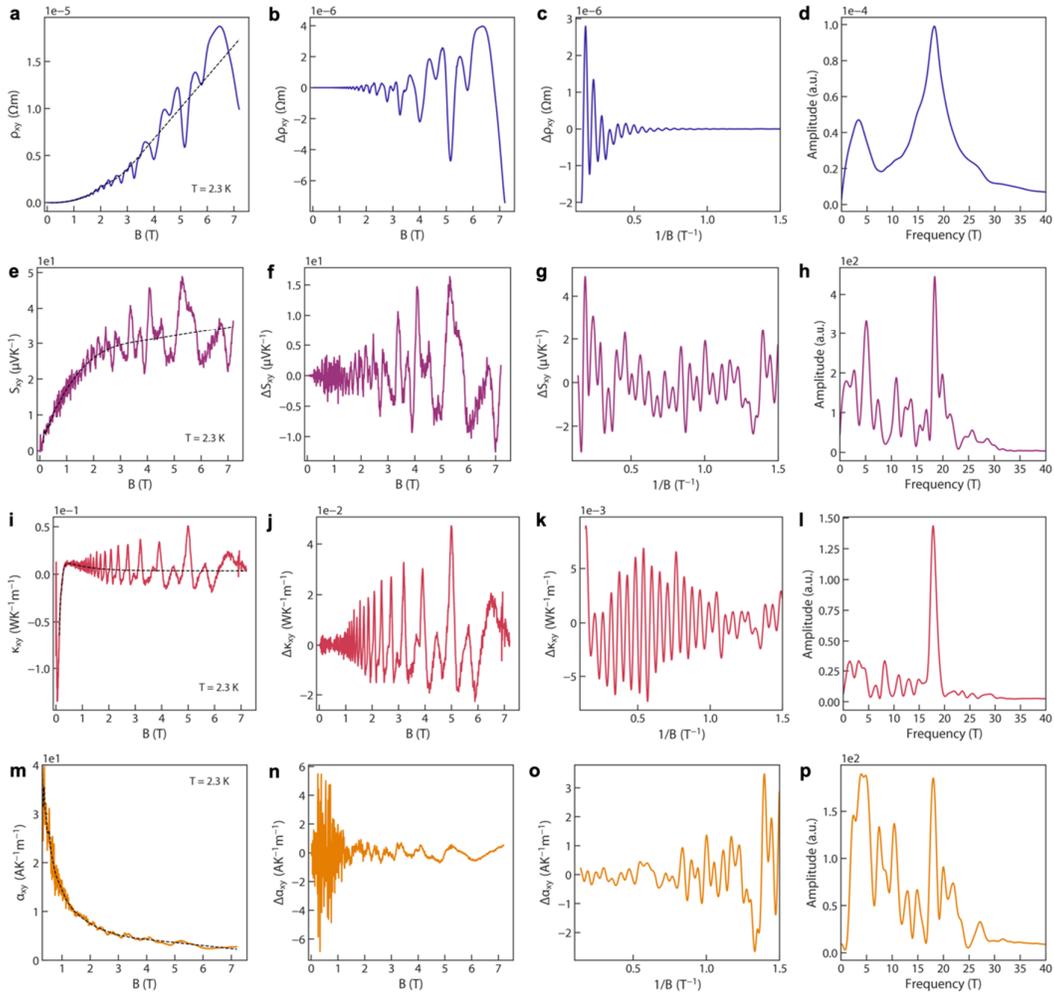

Figure S10. **Quantum oscillation of transverse thermoelectric properties at base temperature. a-d** Transverse resistivity $\rho_{xy}$. **a** $\rho_{xy}$ vs $B$, **b** the background-subtracted part $\Delta\rho_{xy}$ vs $B$, **c** $\Delta\rho_{xy}$ vs $1/B$, and **d** the Fourier transform of **c**. **e-h** Transverse thermopower $S_{xy}$. **e** $S_{xy}$ vs $B$, **f** the background-subtracted part $\Delta S_{xy}$ vs $B$, **g** $\Delta S_{xy}$ vs $1/B$, and **h** the Fourier transform of **g**. **i-l** Transverse thermal conductivity $\kappa_{xy}$. **i** $\kappa_{xy}$ vs $B$, **j** the background-subtracted part $\Delta\kappa_{xy}$ vs $B$, **k** $\Delta\kappa_{xy}$ vs $1/B$, and **l** the Fourier transform of **k**. **m-p** Transverse thermoelectric conductivity $\alpha_{xy}$. **m** $\alpha_{xy}$ vs $B$, **n** the background-subtracted part $\Delta\alpha_{xy}$ vs $B$, **o** $\Delta\alpha_{xy}$ vs $1/B$, and **p** the Fourier transform of **o**.

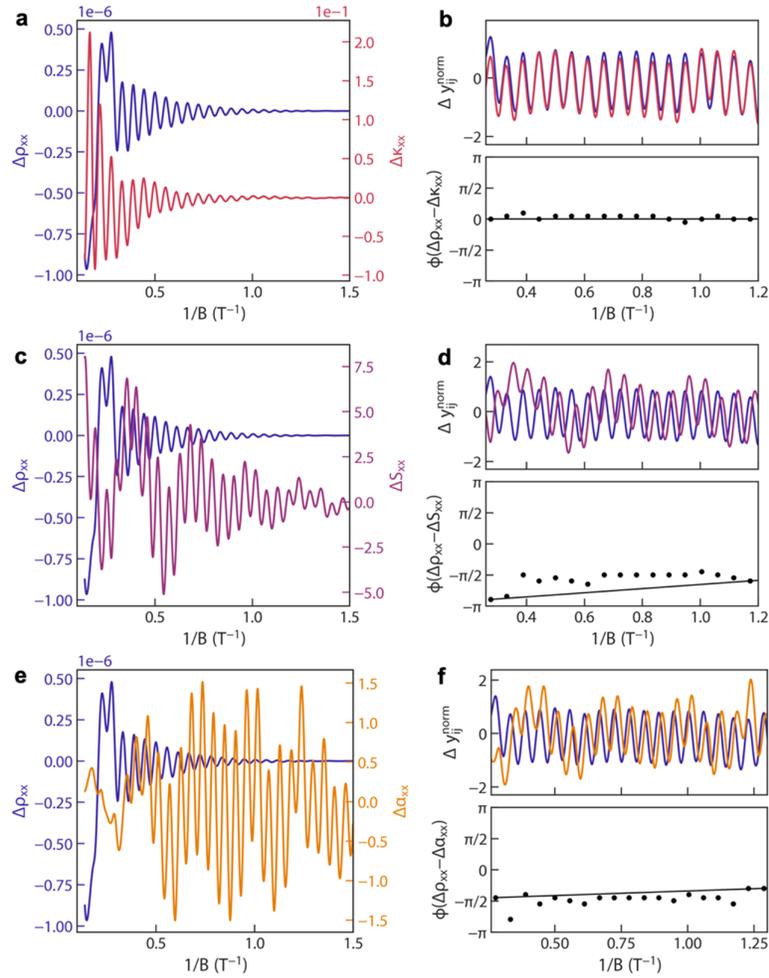

Figure S11. **Phase relations of longitudinal thermoelectric properties. a-b** Resistivity vs thermopower. **a** Quantum oscillation of background-subtracted longitudinal resistivity $\Delta\rho_{xx}$ against the background-subtracted longitudinal thermopower $\Delta S_{xx}$, and **b** the corresponding amplitude-normalized curves highlighting the phase relation between $\Delta\rho_{xx}$ and $\Delta S_{xx}$ (upper figure), and the phase difference between $\Delta\rho_{xx}$ and $\Delta S_{xx}$ (lower figure) as a function of $1/B$. **c-d** Resistivity vs thermal conductivity. **c** Quantum oscillation of background-subtracted longitudinal resistivity $\Delta\rho_{xx}$ against the background-subtracted longitudinal thermal conductivity $\Delta\kappa_{xx}$, and **d** the corresponding amplitude-normalized curves highlighting the phase relation between $\Delta\rho_{xx}$ and $\Delta\kappa_{xx}$ (upper figure), and the phase difference between $\Delta\rho_{xx}$

and $\Delta\kappa_{xx}$ (lower figure) as a function of $1/B$. **e-f** Resistivity vs thermoelectric conductivity. **e** Quantum oscillation of background-subtracted longitudinal resistivity $\Delta\rho_{xx}$ against the background-subtracted longitudinal thermoelectric conductivity $\Delta\alpha_{xx}$, and **f** the corresponding amplitude-normalized curves highlighting the phase relation between $\Delta\rho_{xx}$ and $\Delta\alpha_{xx}$ (upper figure), and the phase difference between $\Delta\rho_{xx}$ and $\Delta\alpha_{xx}$ (lower figure) as a function of $1/B$.

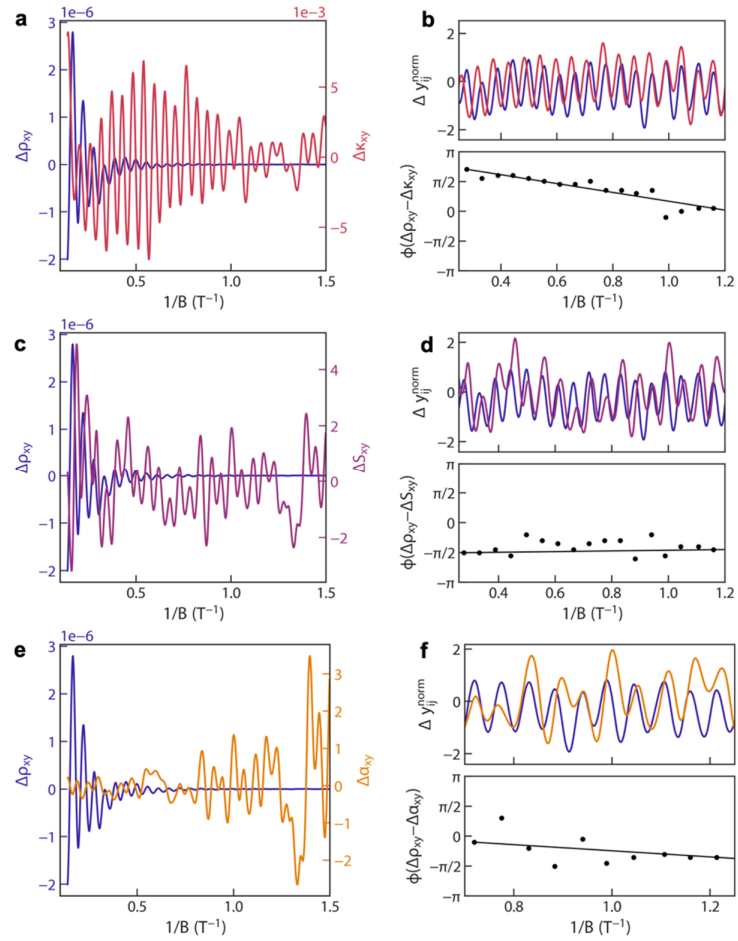

Figure S12. **Phase relations of transverse thermoelectric properties. a-b** Resistivity vs thermopower. **a** Quantum oscillation of background-subtracted transverse resistivity $\Delta\rho_{xy}$ against the background-subtracted transverse thermopower $\Delta S_{xy}$, and **b** the corresponding

amplitude-normalized curves highlighting the phase relation between $\Delta\rho_{xy}$ and $\Delta S_{xy}$ (upper figure), and the phase difference between $\Delta\rho_{xy}$ and $\Delta S_{xy}$ (lower figure) as a function of $1/B$. **c-d** Resistivity vs thermal conductivity. **c** Quantum oscillation of background-subtracted transverse resistivity $\Delta\rho_{xy}$ against the background-subtracted transverse thermal conductivity $\Delta\kappa_{xy}$, and **d** the corresponding amplitude-normalized curves highlighting the phase relation between $\Delta\rho_{xy}$ and $\Delta\kappa_{xy}$ (upper figure), and the phase difference between $\Delta\rho_{xy}$ and $\Delta\kappa_{xy}$ (lower figure) as a function of $1/B$. **e-f** Resistivity vs thermoelectric conductivity. **e** Quantum oscillation of background-subtracted transverse resistivity $\Delta\rho_{xy}$ against the background-subtracted transverse thermoelectric conductivity $\Delta\alpha_{xy}$, and **f** the corresponding amplitude-normalized curves highlighting the phase relation between $\Delta\rho_{xy}$ and $\Delta\alpha_{xy}$ (upper figure), and the phase difference between $\Delta\rho_{xy}$ and $\Delta\alpha_{xy}$ (lower figure) as a function of $1/B$.

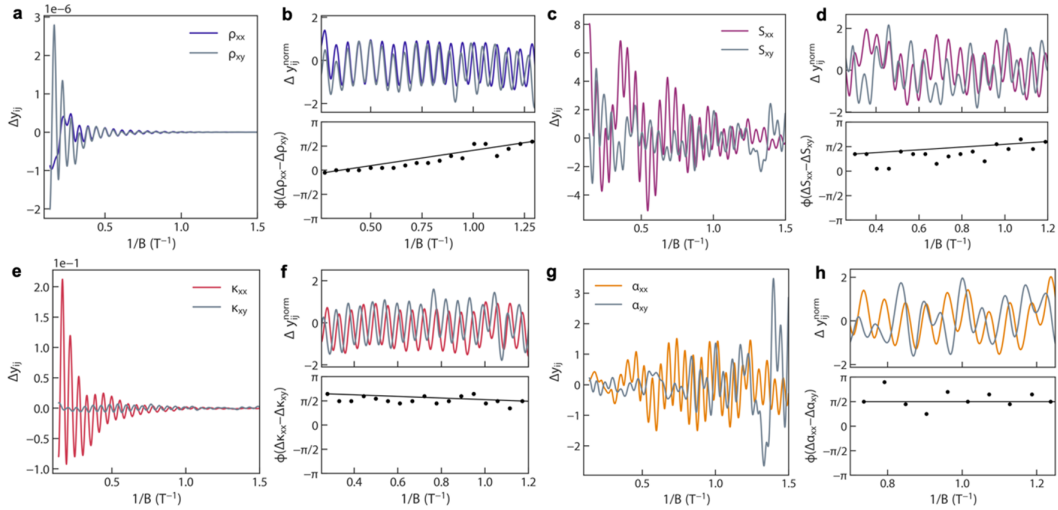

Figure S13. **Phase relations between longitudinal and transverse thermoelectric properties.** **a-b** Phase relation between $\Delta\rho_{xx}$ and $\Delta\rho_{xy}$. **c-d** Phase relation between $\Delta S_{xx}$ and $\Delta S_{xy}$. **e-f** Phase relation between $\Delta\kappa_{xx}$ and $\Delta\kappa_{xy}$. **g-h** Phase relation between $\Delta\alpha_{xx}$ and $\Delta\alpha_{xy}$.

## VII. Thermoelectric Hall Conductivity up to 9T

To validate the quantized thermoelectric Hall effect, particularly the quantized plateau of the thermoelectric Hall coefficient $\alpha_{xy}$ in the high magnetic field limit, we calculated $\alpha_{xy}$ using the following equation:

$$\alpha_{xy} = \frac{-\rho_{xy}S_{xx} + \rho_{xx}S_{xy}}{\rho_{xx}^2 + \rho_{xy}^2} = \frac{\rho_{yx}S_{xx} - \rho_{xx}S_{yx}}{\rho_{xx}^2 + \rho_{yx}^2}. \tag{S10}$$

To obtain $\alpha_{xy}$ as a function of magnetic field for different temperatures, we replotted $S_{xx}$ and $S_{yx}$ from Figures S7e and f, and $\rho_{xx}$ and $\rho_{yx}$ from Figures S8b and c, as functions of magnetic field, as shown in Figures 2d and 3a in the main text and Figures S14a and b in the SI. The resulting $\alpha_{xy}$ calculated with Eq. (S10) is displayed in Figure 3b in the main text.

To extract the values of effective Fermi velocity $v_F^{\text{eff}}$ and chemical potential $\mu$, as well as identify the quantized value of $\alpha_{xy}/T$ approached at very large fields, we fit our low-temperature $\alpha_{xy}$ data up to T=10K using the general expression of $\alpha_{xy}$ in the dissipationless limit [S2] (Eq. (3) of the main text):

$$\alpha_{xy} = \frac{eN_f}{2\pi\hbar} \sum_{n=0}^{\infty}{}' \int_0^{\infty} \frac{dk_z}{\pi} \left[ s\left( \frac{\varepsilon_n^0(k_z) - \mu}{k_B T} \right) + s\left( \frac{\varepsilon_n^0(k_z) + \mu}{k_B T} \right) \right]. \tag{S11}$$

where the notation $\sum_{n=0}^{\infty}{}'$ is used to mean that an extra factor of 1/2 multiplies the n=0 term of the sum; $N_f$ equals the number of Weyl points, and $\varepsilon_n^0(k_z)$ denote the Landau level energies:

$$\varepsilon_n^0(k_z) = \text{sgn(n)} v_F \sqrt{2e\hbar B |n| + \hbar^2 k_z^2} \ . \tag{S12}$$

and $v_F$ is treated as $v_F^{\text{eff}}$. The function $s(x)$ is the entropy per carrier, given by

$$s(x) = -k_B \Big[ n_F(x) \ln n_F(x) + \big( 1 - n_F(x) \big) \ln \big( 1 - n_F(x) \big) \Big]. \tag{S13}$$

where $n_F(x) = \big( 1 + e^{\beta x} \big)^{-1}$ is the Fermi-Dirac distribution. The data to be fitted using Eq. (S11) is shown in Figure S14c, and we extrapolate the fitted function to even larger magnetic fields, revealing we are near the onset of the quantized limit. The value of $\alpha_{xy} / T$ approached in this limit is $\sim 0.4 \text{AK}^{-2} \text{m}^{-1}$. The corresponding fitted parameters are given in Figures S14d and e.

To verify this fit, we additionally fit our low-temperature data up to T=50K using the expression for $\alpha_{xy} / T$ which also includes a finite scattering time $\tau$ and is thus a more expressive form for data with weak scattering present[S2]:

$$\alpha_{xy} = \frac{N_f}{18} \frac{e^2 k_B^2 T v_F \tau^2 B}{\hbar^3} \frac{1 + 3\omega_c^2(E_F)\tau^2}{\big( 1 + \omega_c^2(E_F)\tau^2 \big)^2}. \tag{S14}$$

where the cyclotron frequency $\omega_c$ is given by

$$\omega_c(\varepsilon) = \frac{eB v_F^2}{\varepsilon}. \tag{S15}$$

and once more, $v_F$ is treated as $v_F^{\text{eff}}$. This fit is shown in Figure S14f with the corresponding fitted parameters shown in Figures S14g and h, which are in good agreement with those of the previous fit.

Similarly, we fit our high-temperature data, T>50K, in the limit of weak scattering using

$$\alpha_{xy} = \frac{N_f e^2 k_B^2 T v_F \tau^2 B}{6\pi^2 \hbar^3} \int_{-\infty}^{+\infty} dx \frac{x^4 e^x}{(1+e^x)^2} \frac{1}{x^2 + \omega_c^2 (k_B T)\tau^2}. \qquad (S16)$$

which is shown in Figure S14i with corresponding fitted parameters plotted in Figures S14j and k.

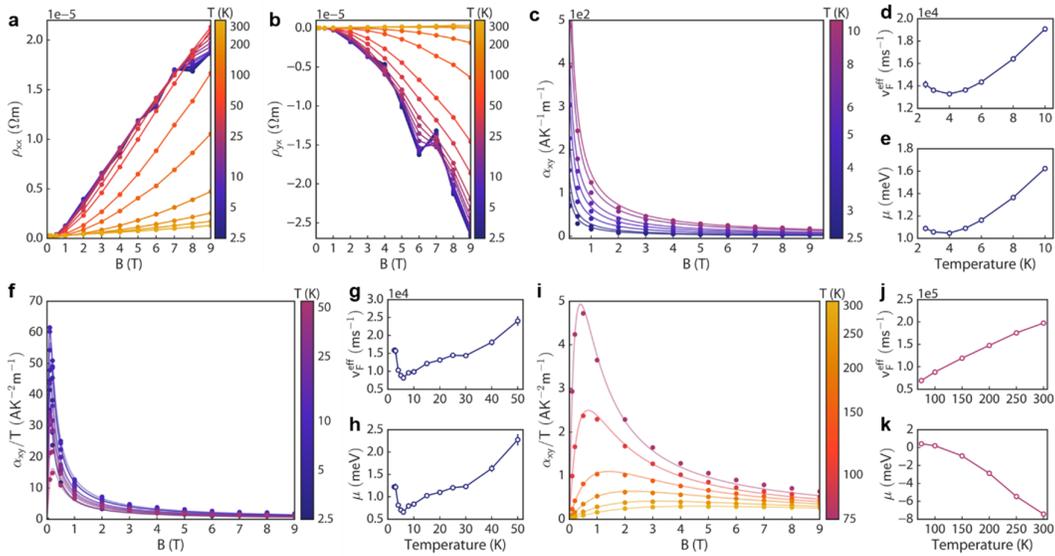

Figure S14. Longitudinal and transverse resistivities **a** $\rho_{xx}$ and **b** $\rho_{yx}$ as functions of magnetic field at different temperatures. **c** Thermoelectric Hall conductivity $\alpha_{xy}$ as a function of magnetic field at different temperatures. The solids lines are fitted curves using Eq. (S11) (low-temperature dissipationless limit), shown as solid lines. **d-e** Effective Fermi velocity $v_F^{\text{eff}}$ and chemical potential $\mu$ obtained from the fitting in **c** using on Eq. (S11). **f** $\alpha_{xy}/T$ fitted using Eq. (S14), where the corresponding fitting parameters are shown in **g** and **h**. At higher temperature, **i** the $\alpha_{xy}/T$ is fitted with Eq. (S16), and the corresponding fitted parameters are shown in **j** and **k**. It can be seen that there is a general quantitative agreement using different fitting equations.

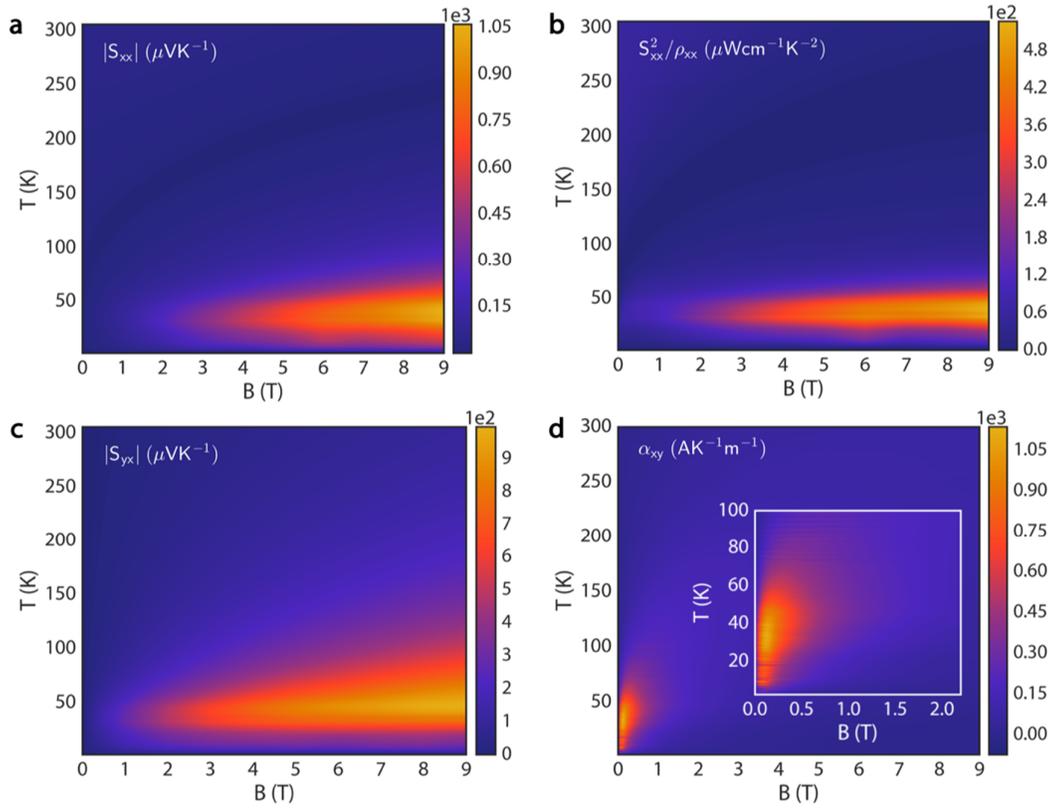

Figure S15. The 2D contour plots of **a** $\left|S_{xx}\right|$, **b** $S_{xx}^2/\rho_{xx}$, **c** $\left|S_{yx}\right|$, **d** $\alpha_{xy}$ showing comprehensive data collection from $B$=0T to 9T, and from $T$=2K to 300K.

## VIII. Low-temperature Thermoelectric Measurements up to 14T

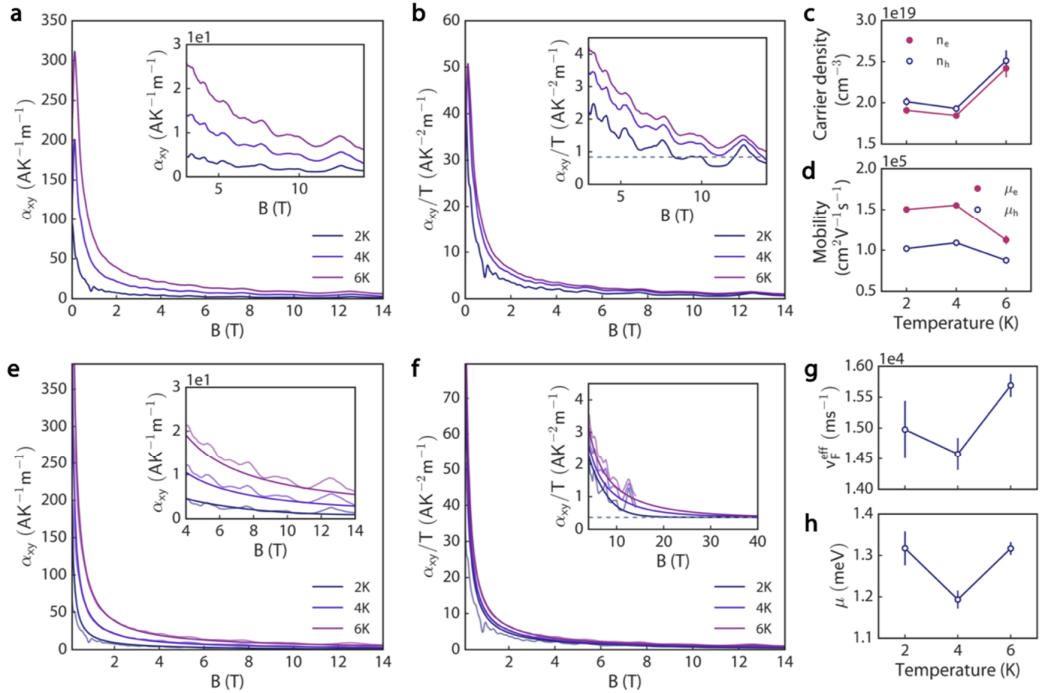

Figure S16. The thermoelectric Hall conductivity **a** $\alpha_{xy}$ and **b** the ratio $\alpha_{xy}/T$ at $T$=2K, 4K and 6K up to 14T. We can see a clear flattening trend that persists beyond 9T whereby the three different temperature curves collapse into one. The **c** carried density and the **d** mobility obtained by fitting. **e-f** Identical $\alpha_{xy}$ and $\alpha_{xy}/T$ data, overlaid on top of the fitting; the universality can be seen by extending to high magnetic field, resulting in the universal value $\alpha_{xy}/T = 0.37 \mathrm{AK^{-2}m^{-1}}$, consistent with the separate 9T data. **g-h** The effective Fermi velocity and chemical potential are also in excellent agreement with the 9T data.

## IX. Dominant Thermoelectric Hall Contribution to Longitudinal Thermoelectric Performance at Low Temperatures

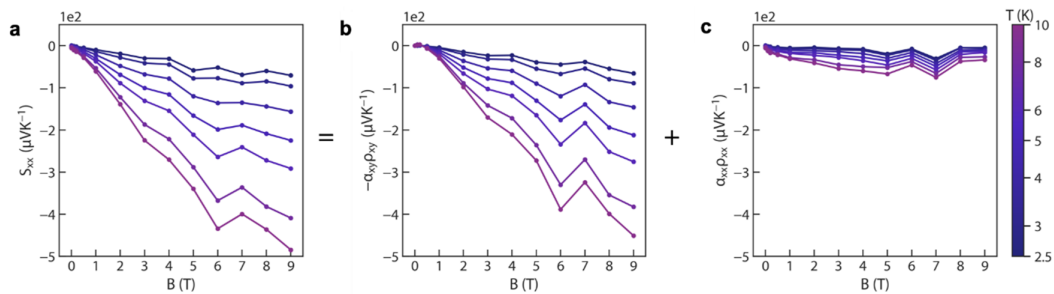

Figure S17. **Dominant contribution of longitudinal thermopower** $S_{xx}$ **from the transverse thermoelectric Hall conductivity** $\alpha_{xy}$ **at low temperatures. a** Total $S_{xx}$ up to 10K as a function of magnetic field separated into **b** a transverse contribution $-\alpha_{xy}\rho_{xy}$ and **c** a longitudinal contribution $+\alpha_{xx}\rho_{xx}$ **c**. All results show that the transverse component dominantly contributes over 90% of the longitudinal thermopower value at low temperatures.

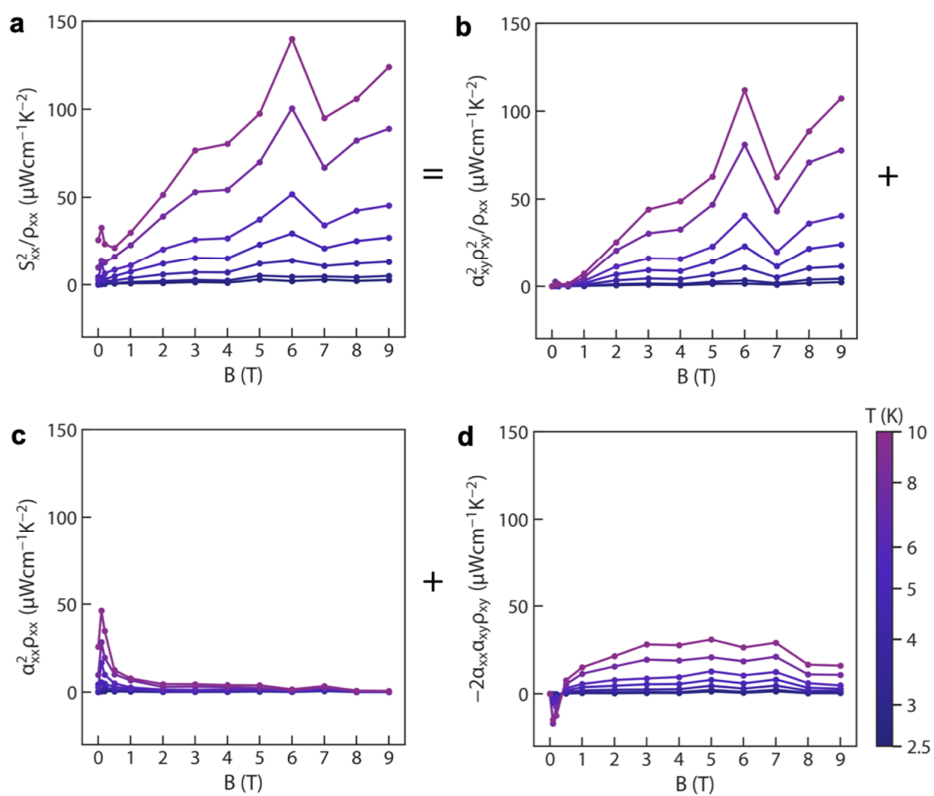

Figure S18. **Dominant contribution of longitudinal power factor** $S_{xx}^2 / \rho_{xx}$ **from the transverse thermoelectric Hall conductivity** $\alpha_{xy}$ **at low temperatures. a** Total $S_{xx}^2 / \rho_{xx}$ at low temperatures as a function of magnetic field separated into **b** a transverse component $\alpha_{xy}^2 \rho_{xy}^2 / \rho_{xx}$, **c** a longitudinal contribution $+\alpha_{xx}^2 \rho_{xx}$, and **d** a cross term contribution $-2\alpha_{xy}\alpha_{xx}\rho_{xy}$.

## X. X-Ray and Neutron Scattering Measurement Details

Inelastic neutron scattering measurements were performed on the HB1 triple-axis spectrometer at the High-Flux Isotope Reactor at the Oak Ridge National Laboratory. We used a fixed $E_f$ = 14.7 meV with 48'−40'−40'−120' collimation and Pyrolytic Graphite filters to eliminate higher-harmonic neutrons. Measurements were performed using closed-cycle refrigerators between room temperature and the base temperature 4 K. Inelastic X-ray scattering was performed on the high-energy resolution inelastic x-ray (HERIX) instrument at sector 3-ID beamline of the Advanced Photon Source, Argonne National Laboratory with incident beam energy of 21.657 keV (λ=0.5725Å) and an overall energy resolution of 2.1 meV [S4-6]. Incident beam focused on the sample using toroidal and KB mirror system. FWHM of beam size at sample position was 20 × 20 μm² (V × H). The spectrometer was functioning in the horizontal scattering geometry with a horizontally polarized radiation. The scattered beam was analyzed by a diced and spherically curved silicon (18 6 0) analyzers working at backscattering angle. The basic principles of such instrumentations are discussed elsewhere [S7,8].

Measurements of the phonon modes along high-symmetry lines in the Brillouin zone of TaP were performed using both inelastic x-ray scattering and inelastic neutron scattering. Selected raw intensity spectra along high symmetry direction Γ to Σ are shown in Figure S19 using x-rays (left) and neutrons (right). The spectra were analyzed by a damped harmonic oscillator (DHO) model convoluted with the experimental resolution function to yield the energy and intensity of each mode. These were used to generate a phonon dispersion relation, which can be seen in Figure 5c in the main text, along high symmetry line Z-Γ-Σ. These experimental results serve as a consistency check to support the ab initio calculations performed for the thermal conductivity used in the main text and displayed in Figure 5b.

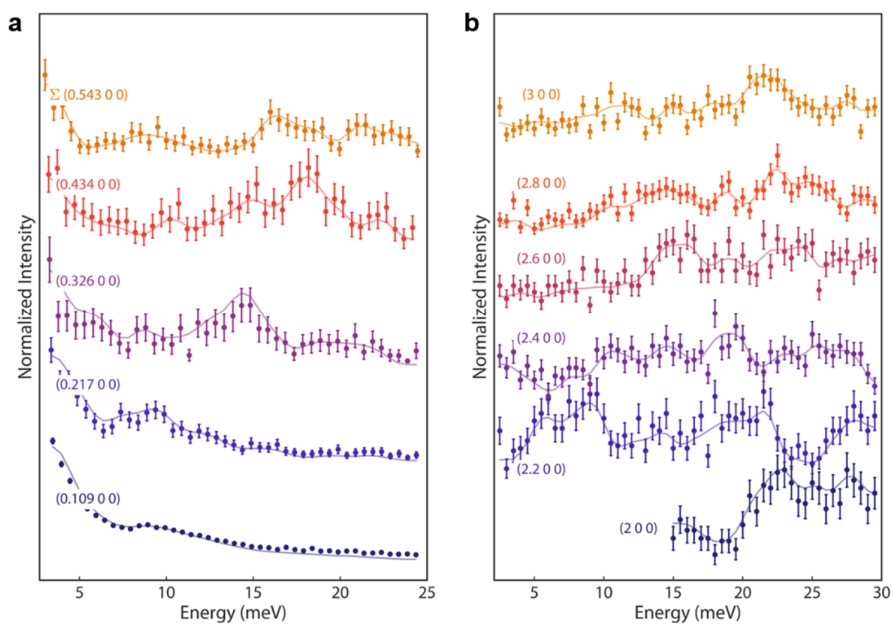

Figure S19. **a** X-ray and **b** neutron inelastic scattering measurements along the high symmetry direction Γ-Σ. The faint solid lines are a guide for the eye.

## XI. Separation of Phonon and Electron Contributions to Thermal Conductivity

To check the compliance or violation of the Wiedemann-Franz law, the phononic and electronic contributions to thermal conductivity need to be separated.

To separate the phononic and electronic contributions, we fit $\kappa_{xx}$ versus $B$ curves with the following empirical equation:

$$\kappa_{xx}(T,B) = \kappa_{xx}^{ph}(T) + \frac{\kappa_{xx}^{e}(T, B = 0\text{T})}{1 + \beta_e(T)B^m}. \tag{S17}$$

where $\beta_e(T)$ is proportional to the zero-field electronic mean free path of electrons, and $m$ is related to the nature of the electron scattering[S9,10].

Figures S20b-d shows the gradual suppression of $\kappa_{xx}$ at high magnetic fields at typical temperatures 100K, 200K and 300K. We can see at 100K, $\kappa_{xx}$ forms a plateau above 4T, indicating that the electronic thermal conductivity is almost completely suppressed, while at 200K and 300K, the suppression is still in an intermediate state. All the $\kappa_{xx}$ versus $B$ curves can be fitted well with Eq. (S17) and the fitting process for different temperatures successfully achieves the separation of the phononic and electronic contributions to thermal conductivity. The resulting phononic and electronic thermal conductivities are discussed in detail in the main text. Here we stress that the fitting parameter $\beta_e(T)$ obtained from the fitting shows a typical behavior of thermally elevated electron-phonon scattering, as shown in Figure S20a, and the fitting parameter $n$ fluctuates around 1.35, indicating its constant nature which implies that our fitting process is reasonable.

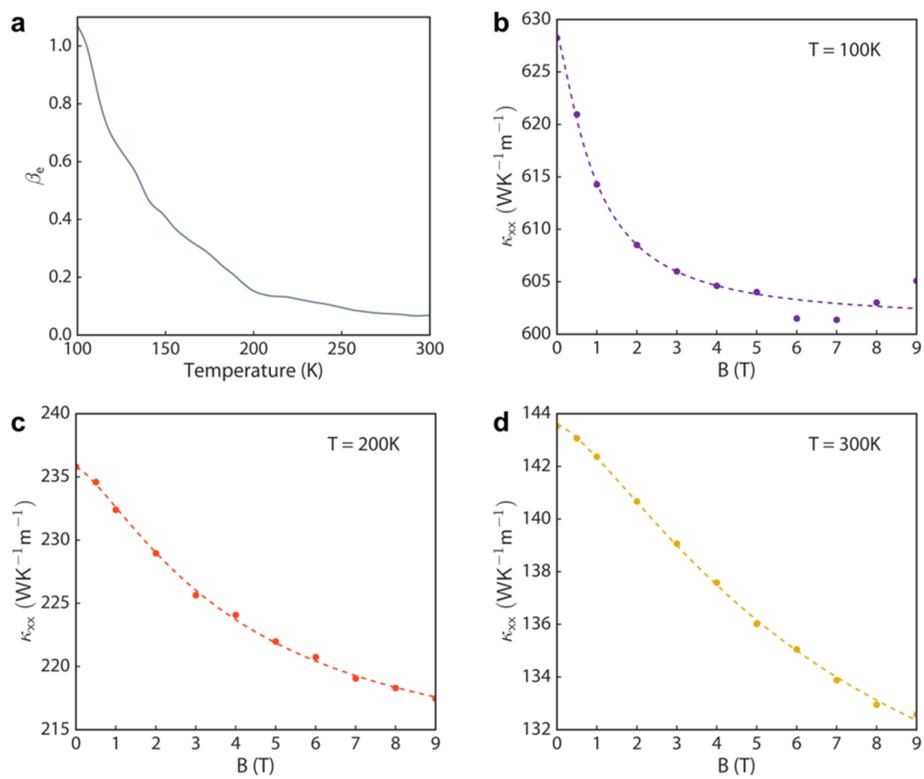

Figure S20. **a** The fitting parameter $\beta_e(T)$ as a function of temperature. The fitting for the $\kappa_{xx}$ versus $B$ curves at **b** 100K, **c** 200K and **d** 300K.

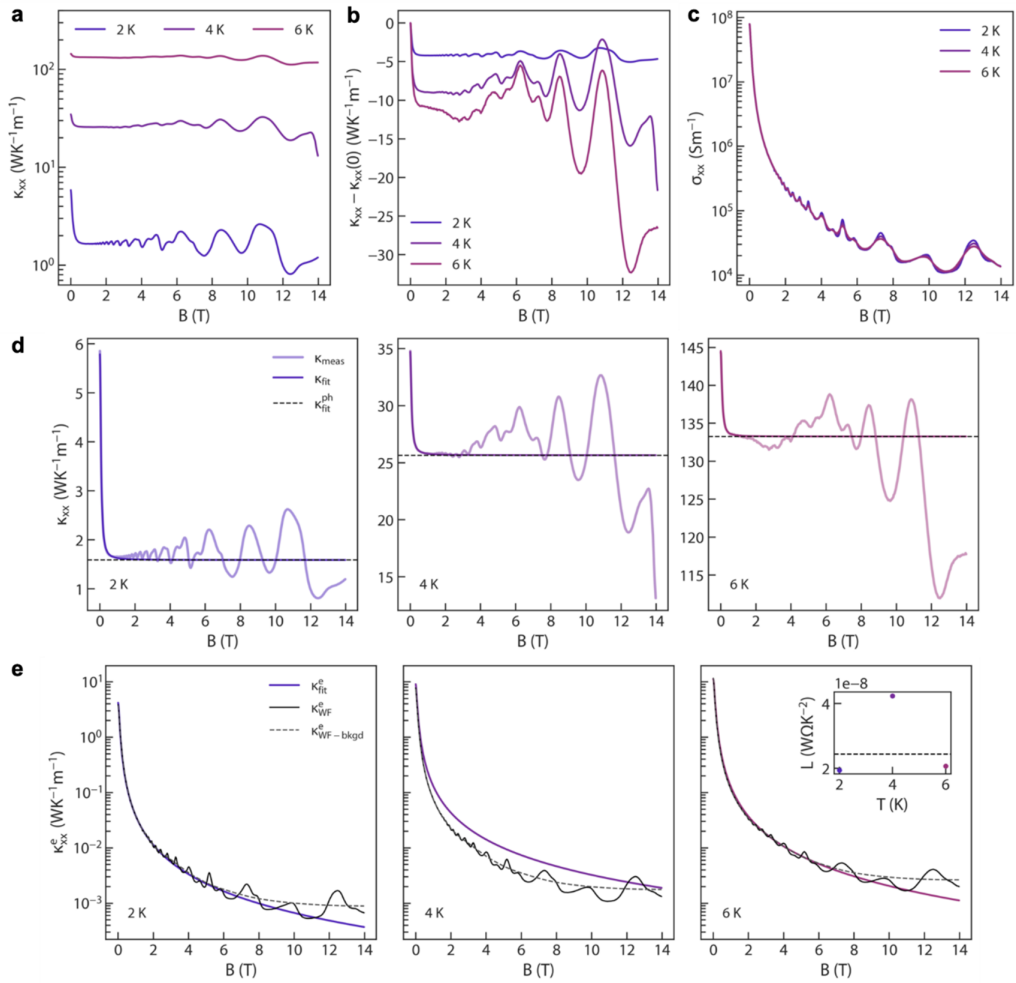

Figure S21. **Wiedemann-Franz law in the low-temperature regime up to *B*=14T. a** Longitudinal thermal conductivity $\kappa_{xx}$ at 2K, 4K, and 6K as a function of B. **b** Longitudinal thermal conductivity relative to its value at zero field. **c** Longitudinal conductivity $\sigma_{xx}$ at 2K, 4K, and 6K as a function of B. **d** The result of fitting the empirical equation S17 to $\kappa_{xx}$ at 2K, 4K, and 6K (left to right). The fitted value of the phonon contribution to $\kappa_{xx}$ is indicated in each plot by the dashed black line. **e** Comparison of the electronic contribution to $\kappa_{xx}$ obtained by fitting and by direct calculation using the Wiedemann-Franz law and measured $\sigma_{xx}$, showing good agreement at low temperature. Due to the presence of quantum oscillations at high field, a calculation using a smooth background of $\sigma_{xx}$ is also performed (dashed line).

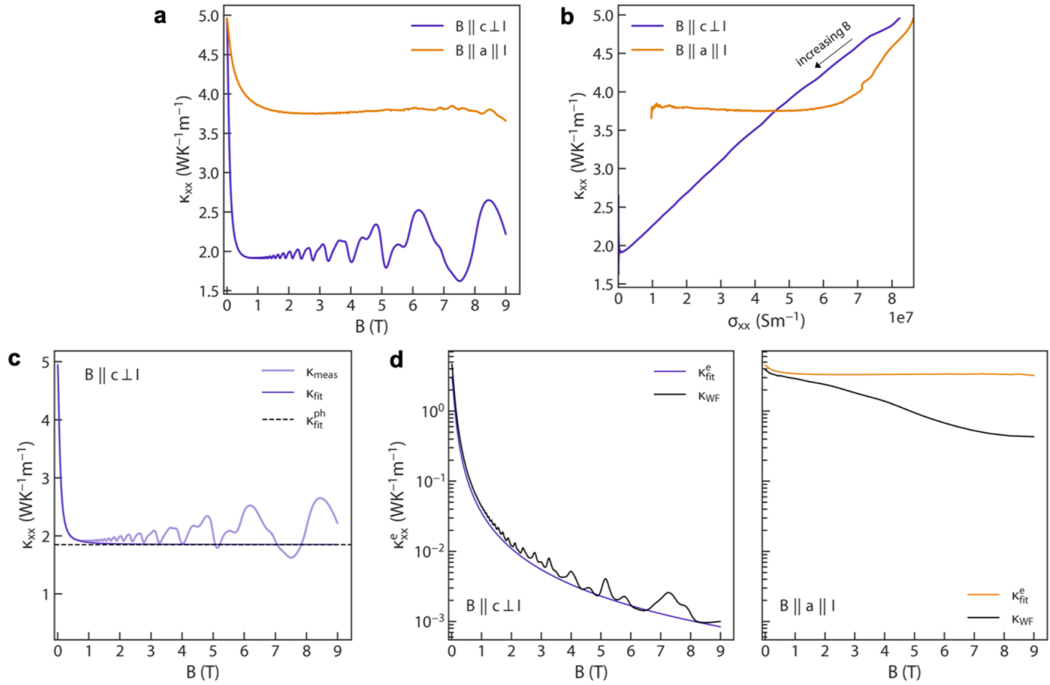

Figure S22. **Comparison between configurations with B ∥ c and B ∥ a. a.** Longitudinal thermal conductivity $\kappa_{xx}$ at the base temperature 2K as a function of B, applied either parallel to the c-axis direction (perpendicular to the current I direction) or parallel to the a-axis direction (parallel to the current direction). **b** $\kappa_{xx}$ versus the longitudinal conductivity $\sigma_{xx}$ under the two different configurations. **c** The result of fitting the empirical equation S17 to $\kappa_{xx}$ in the B ∥ c configuration only. The fitted value of the phonon contribution to $\kappa_{xx}$ is indicated by the dashed black line. **d** Comparison of the electronic contribution to $\kappa_{xx}$ obtained by fitting and by direct calculation using the Wiedemann-Franz law and measured $\sigma_{xx}$ for both configurations. To obtain the fitted value of the electronic contribution in the case of B ∥ a, the phonon contribution obtained by fitting B ∥ c was adjusted by the ratio of sound velocities along the a- and c-axis directions, which were estimated from the calculated phonon dispersion (see Fig. 4c of the main text). The results show good agreement with the Wiedemann-Franz law for the case B ∥ c, but a departure by approximately one order of magnitude at the maximum field for the B ∥ a configuration.

### XII. Computational Details

All the ab initio calculations are performed by Vienna Ab Initio Package (VASP)[S11,12] with projector-augmented-wave (PAW) pseudopotentials and Perdew-Burke-Ernzerhof (PBE) for exchange-correlation energy functional[S13]. The geometry optimization of the conventional cell was performed with a $6 \times 6 \times 2$ Monkhorst-Pack grid of k-point sampling. The second-order and third-order force constants was calculated using a real space supercell approach with a $3 \times 3 \times 1$ supercell, same as Ref[S14]. The Phonopy package[S15] was used to obtain the second-order force constants. The thirdorder.py and ShengBTE packages[S16] were used to obtain the third-order force constants and relaxing time approximation was used to calculate the thermal conductivity. A cutoff radius of about 0.42 nm was used, which corresponds to including the fifth nearest neighbor when determining the third-order force constants. To get the equilibrium distribution function and scattering rates using the third-order force constants, the first Brillouin zone was sampled with 30×30×10 mesh.